\documentclass[letter,twocolumn,showpacs,aps,prd,preprintnumbers,superscriptaddress]{revtex4}
\usepackage{graphicx}
\usepackage{dcolumn}
\usepackage{amsmath}
\usepackage{epsfig}
\usepackage{color}
\usepackage{rotating}

\usepackage{relsize}
\RequirePackage{xspace}

\def\pep2{PEP-II}
\def\babar{\mbox{\slshape B\kern-0.1em{\smaller A}\kern-0.1em
    B\kern-0.1em{\smaller A\kern-0.2em R}}}
\def\cm   {\ensuremath{{\rm \,cm}}\xspace}
\def\Y#1S{\ensuremath{\Upsilon{(#1S)}}\xspace}
\def\FourS {\Y4S}
\def\Kz    {\ensuremath{K^0}\xspace}
\def\KS    {\ensuremath{K^0_{\scriptscriptstyle S}}\xspace} 
\def\KL    {\ensuremath{K^0_{\scriptscriptstyle L}}\xspace}

\def\epem       {\ensuremath{e^+e^-}\xspace}

\def\ccbar {\ensuremath{c\overline c}\xspace}

\def\pip   {\ensuremath{\pi^+}\xspace}
\def\pim   {\ensuremath{\pi^-}\xspace}

\def\Kp    {\ensuremath{K^+}\xspace}
\def\Km    {\ensuremath{K^-}\xspace}

\def\Dbar    {\kern 0.2em\overline{\kern -0.2em D}{}\xspace}

\def\Dz      {\ensuremath{D^0}\xspace}
\def\Dzb     {\ensuremath{\Dbar^0}\xspace}
\def\DzDzb   {\ensuremath{\Dz {\kern -0.16em \Dzb}}\xspace}
\def\Dp      {\ensuremath{D^+}\xspace}
\def\Dm      {\ensuremath{D^-}\xspace}
\def\DpDm    {\ensuremath{\Dp {\kern -0.16em \Dm}}\xspace}
\def\Dstar   {\ensuremath{D^*}\xspace}

\def\Dstarz  {\ensuremath{D^{*0}}\xspace}

\def\Dstarp  {\ensuremath{D^{*+}}\xspace}

\def\Dsp     {\ensuremath{D^+_s}\xspace}
\def\DsJs    {\ensuremath{D^{*}_{s0}(2317)^{+}}\xspace}
\def\DsJ     {\ensuremath{D_{s1}(2460)^{+}}\xspace}

\def\Dsop    {\ensuremath{D^+_{s1}}\xspace}

\def\Dsom    {\ensuremath{D^-_{s1}}\xspace}
\def\Dsopnum {\ensuremath{D_{s1}(2536)^{+}}\xspace}

\def\dmds  {\ensuremath{\Delta m(\Dsop)}\xspace}
\def\dmdsz {\ensuremath{\Delta m(\Dsop)_{0}}\xspace}
\def\gds {\ensuremath{\Gamma(\Dsop)}\xspace}
\def\pstar {\ensuremath{p^{\ast}(\Dsop)}\xspace}
\def\delm {\ensuremath{\Delta_{\Delta m}}\xspace}
\def\delg {\ensuremath{\Delta_{\Gamma}}\xspace}
\def\mum  {\ensuremath{\,\mu\rm m}\xspace}

\newcommand{\tev}{\ensuremath{\mathrm{\,Te\kern -0.1em V}}\xspace}
\newcommand{\gev}{\ensuremath{\mathrm{\,Ge\kern -0.1em V}}\xspace}
\newcommand{\mev}{\ensuremath{\mathrm{\,Me\kern -0.1em V}}\xspace}
\newcommand{\kev}{\ensuremath{\mathrm{\,ke\kern -0.1em V}}\xspace}
\newcommand{\ev}{\ensuremath{\mathrm{\,e\kern -0.1em V}}\xspace}
\newcommand{\kevc}{\ensuremath{{\mathrm{\,ke\kern -0.1em V\!/}c}}\xspace}
\newcommand{\gevc}{\ensuremath{{\mathrm{\,Ge\kern -0.1em V\!/}c}}\xspace}
\newcommand{\mevc}{\ensuremath{{\mathrm{\,Me\kern -0.1em V\!/}c}}\xspace}
\newcommand{\gevcc}{\ensuremath{{\mathrm{\,Ge\kern -0.1em V\!/}c^2}}\xspace}
\newcommand{\mevcc}{\ensuremath{{\mathrm{\,Me\kern -0.1em V\!/}c^2}}\xspace}
\newcommand{\kevcc}{\ensuremath{{\mathrm{\,ke\kern -0.1em V\!/}c^2}}\xspace}

\def\invfb   {\ensuremath{\mbox{\,fb}^{-1}}\xspace}

\newcommand{\jprlBase}       {Phys.\ Rev.\ Lett.\xspace}
\newcommand{\jprBase}        {Phys.\ Rev.\xspace}
\newcommand{\jplBase}        {Phys.\ Lett.\xspace}
\newcommand{\nimBaseA}       {Nucl.\ Instrum.\ Methods Phys.\ Res., Sect.\ A\xspace}

\newcommand{\npBase}         {Nucl.\ Phys.\xspace}

\newcommand{\app}       [1]  {{Acta Phys.\ Polon.\ {\bf #1}}}

\newcommand{\cpc}       [1]  {{Comput.\ Phys.\ Commun.\ {\bf #1}}}

\newcommand{\jpg}       [1]  {{J.\ Phys.\ {\bf G{\bf #1}}}}

\newcommand{\mpl}       [1]  {{Mod.\ Phys.\ Lett.\ {\bf #1}}}

\newcommand{\nima}      [1]  {\nimBaseA~{\bf #1}}

\newcommand{\npb}       [1]  {\npBase\ B~{\bf #1}}

\newcommand{\plb}       [1]  {\jplBase\ B~{\bf #1}}
\newcommand{\prep}      [1]  {{Phys.\ Rep.\ {\bf #1}}}
\newcommand{\jprl}      [1]  {\jprlBase\ {\bf #1}}

\newcommand{\jprd}      [1]  {\jprBase\ D~{\bf #1}}

\newcommand{\BABARPubYear}    {10}
\newcommand{\BABARPubNumber}  {031}

\newcommand{\SLACPubNumber} {14400}

\def\figurebox#1#2#3{
    \def\arg{#3}
    \ifx\arg\empty
    {\hfill\vbox{\hsize#2\hrule\hbox to #2{\vrule\hfill\vbox to #1{\hsize#2\vfill}\vrule}\hrule}\hfill}
    \else
    {\hfill\epsfbox{#3}\hfill}
    \fi}

\begin{document}

\preprint{\babar-PUB-\BABARPubYear/\BABARPubNumber}
\preprint{SLAC-PUB-\SLACPubNumber}

\title{\bf Measurement of the mass and width of the {\boldmath \Dsopnum} meson}

\author{J.~P.~Lees}
\author{V.~Poireau}
\author{E.~Prencipe}
\author{V.~Tisserand}
\affiliation{Laboratoire d'Annecy-le-Vieux de Physique des Particules (LAPP), Universit\'e de Savoie, CNRS/IN2P3,  F-74941 Annecy-Le-Vieux, France}
\author{J.~Garra~Tico}
\author{E.~Grauges}
\affiliation{Universitat de Barcelona, Facultat de Fisica, Departament ECM, E-08028 Barcelona, Spain }
\author{M.~Martinelli$^{ab}$}
\author{D.~A.~Milanes$^{ab}$ }
\author{A.~Palano$^{ab}$ }
\author{M.~Pappagallo$^{ab}$ }
\affiliation{INFN Sezione di Bari$^{a}$; Dipartimento di Fisica, Universit\`a di Bari$^{b}$, I-70126 Bari, Italy }
% ...
\collaboration{The \babar\ Collaboration}
\noaffiliation

\date{\today}

\begin{abstract}
The decay width and mass of the \Dsopnum meson are measured via the decay channel $\Dsop \to \Dstarp\KS$ using $385\invfb$ of data recorded with the \babar~detector in the vicinity of the \FourS resonance at the PEP-II asymmetric-energy electron-positron collider. The result for the decay width is $\gds = 0.92 \pm 0.03\,(\mathrm{stat.}) \pm 0.04\,(\mathrm{syst.})\mev$. For the mass, a value of $m(\Dsop) = 2535.08 \pm 0.01\,(\mathrm{stat.}) \pm 0.15\,(\mathrm{syst.})\mevcc$ is obtained. The mass difference between the \Dsop and the \Dstarp is measured to be $m(\Dsop)-m(\Dstarp) = 524.83 \pm 0.01\,(\mathrm{stat.}) \pm 0.04\,(\mathrm{syst.})\mevcc$, representing a significant improvement compared to the current world average. The unnatural spin-parity assignment for the \Dsop meson is confirmed. 
\end{abstract}

\maketitle

\section{Introduction}
\label{sec:intro}
The theoretical description of \Dsp mesons is problematic because, unlike $D$ mesons, the masses and widths of the \DsJs and \DsJ states~\cite{Au03,Au04,Mi04,Au06,Be03,Kr03} are not in agreement with potential model calculations based on HQET~\cite{Ca03}. Theoretical explanations for the discrepancy invoke $D^{(\ast)}K$ molecules~\cite{Ba03}, chiral partners~\cite{Bd03, No03}, unitarized chiral models~\cite{Bv03, Ko04}, tetraquarks~\cite{Ma05,Te03}, and lattice calculations~\cite{Do03,Bl03}, but a satisfactory description is still lacking (see~\cite{Co04, Sw06} for more details). Improved measurements of the \Dsop meson parameters can lead to a better understanding of these states.\\
\indent In this analysis a precise measurement of the \Dsopnum mass and decay width is performed based on a high statistics data sample~\cite{CC}. The \Dsopnum meson, referred to as the \Dsop below, was first seen in \ccbar-continuum reactions~\cite{Al89}, and more recently in $B$ decays. The current world average mass value published by the Particle Data Group is based on measurements with large statistical and systematic uncertainties: $2535.29 \pm 0.20\mevcc$~\cite{Pd10}; the mass difference between the \Dsop and the \Dstarp meson has been measured to be $525.04 \pm 0.22\mevcc$~\cite{Pd10}. An upper limit on the decay width ($\Gamma<2.3\mev$ at $90\%$ confidence level), and a measurement of the spin-parity of the \Dsop meson ($J^{P}=1^{+}$), have been obtained, but based on low-statistics data samples only~\cite{Pd10,Au08,Al93}. The mixing between the \Dsop meson and the other $J^{P}=1^{+}$ state \DsJ was investigated in Ref.~\cite{Ba08}.  \\
\indent This analysis is based on a data sample corresponding to an integrated luminosity of $349\invfb$ recorded at the \FourS resonance and $36\invfb$ recorded $40\mev$ below that resonance with the \babar~detector at the asymmetric-energy \epem collider PEP-II at the SLAC National Accelerator Laboratory. In this analysis, \Dsop mesons are reconstructed from \ccbar continuum events in the $\Dstarp\KS$ channel; those originating from $B$ decays are rejected. \\
\indent The \babar~detector is described briefly in Sec.~\ref{sec:det}. The principal criteria used in the reconstruction of the $\Dstarp\KS$ mass spectrum and the selection of \Dsop-candidates are discussed in Sec.~\ref{sec:reco}. The relevant Monte Carlo (MC) simulations are described in Sec.~\ref{sec:mc}, while the detector resolution parametrization is considered in Sec.~\ref{sec:res}. Measurements of the mass and total width for the \Dsop state are obtained from a fit to the $\Dstarp\KS$ invariant mass distribution as discussed in Sec.~\ref{sec:datafit}. Decay angle distributions are studied in Sec.~\ref{sec:angdist}, where the implications for the spin-parity of the \Dsop state are also discussed. Sources of systematic uncertainty are described in Sec.~\ref{sec:syst}, and the results of the analysis are summarized in Secs.~\ref{sec:results} and~\ref{sec:summ}.

\section{The \babar~Detector}
\label{sec:det}
The \babar~detector is described in detail elsewhere~\cite{Au02}. Charged particles are detected, and their momenta measured, with a combination of five layers of double-sided silicon microstrip detectors (SVT) and a 40-layer cylindrical drift chamber (DCH), both coaxial with the cryostat of a superconducting solenoidal magnet that produces a magnetic field of $1.5~\mathrm{T}$. Charged particle identification is achieved by measurements of the energy loss $dE/dx$ in the tracking devices and with an internally reflecting, ring-imaging Cherenkov detector. The energy of photons and electrons is measured with a CsI(Tl) electromagnetic calorimeter, covering $90\%$ of the $4\pi$ solid angle in the $\Upsilon(4S)$ rest frame. The instrumented flux return of the magnetic field is used to identify muons and $\KL$'s.

\section{Selection and Reconstruction of events}
\label{sec:reco}
\begin{figure*}
\includegraphics[width=0.3\textwidth]{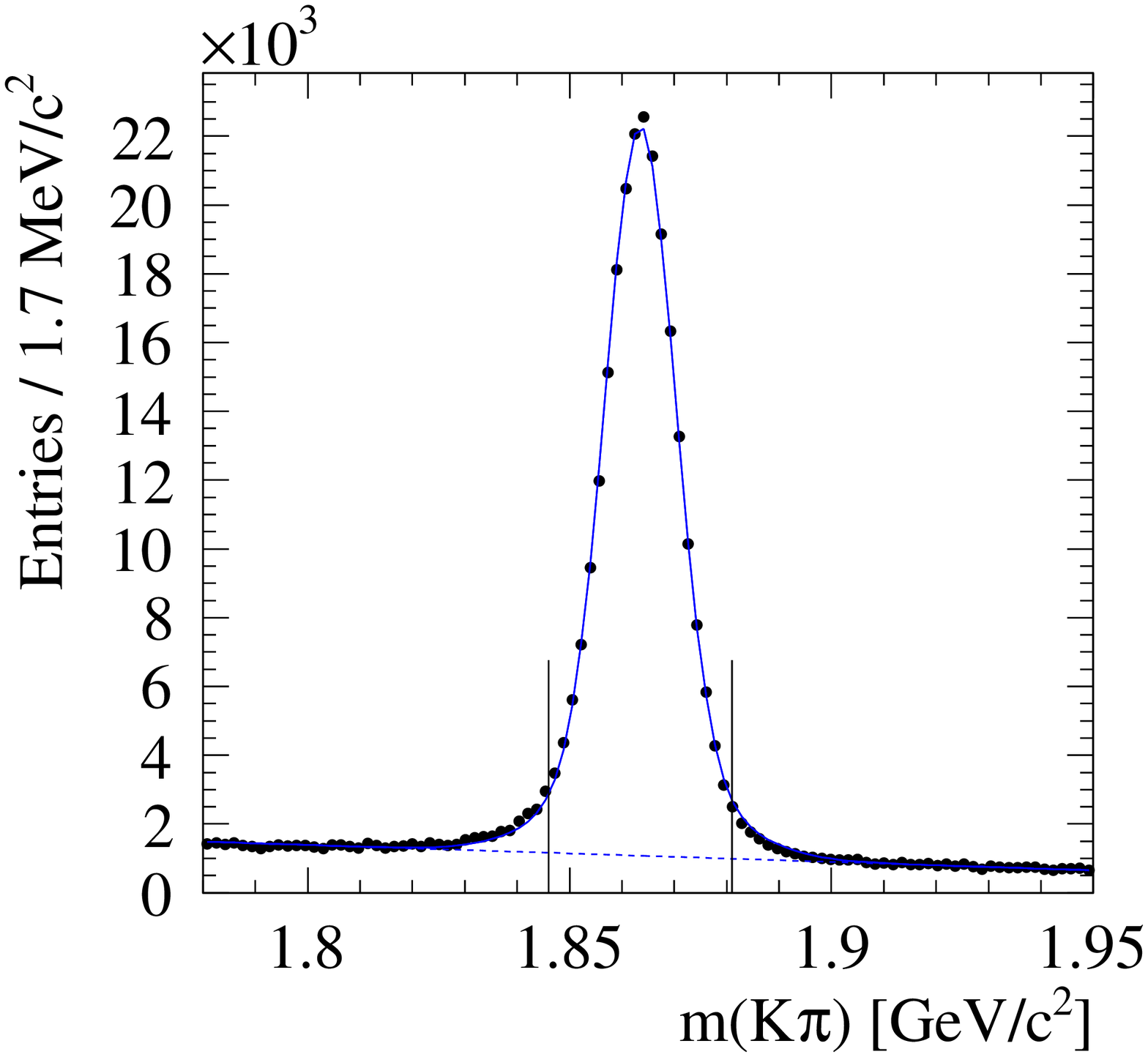}
\includegraphics[width=0.3\textwidth]{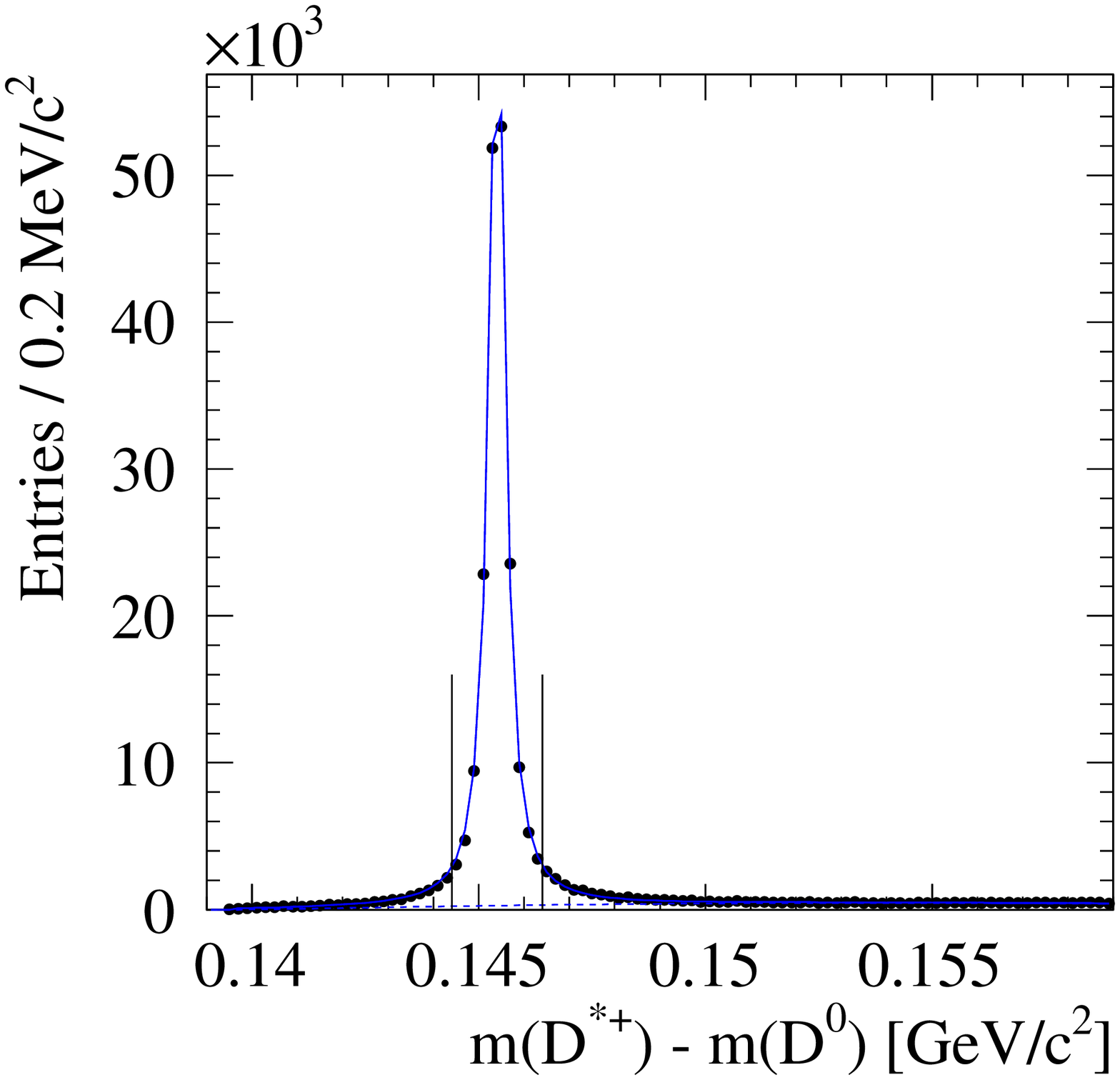}
\includegraphics[width=0.3\textwidth]{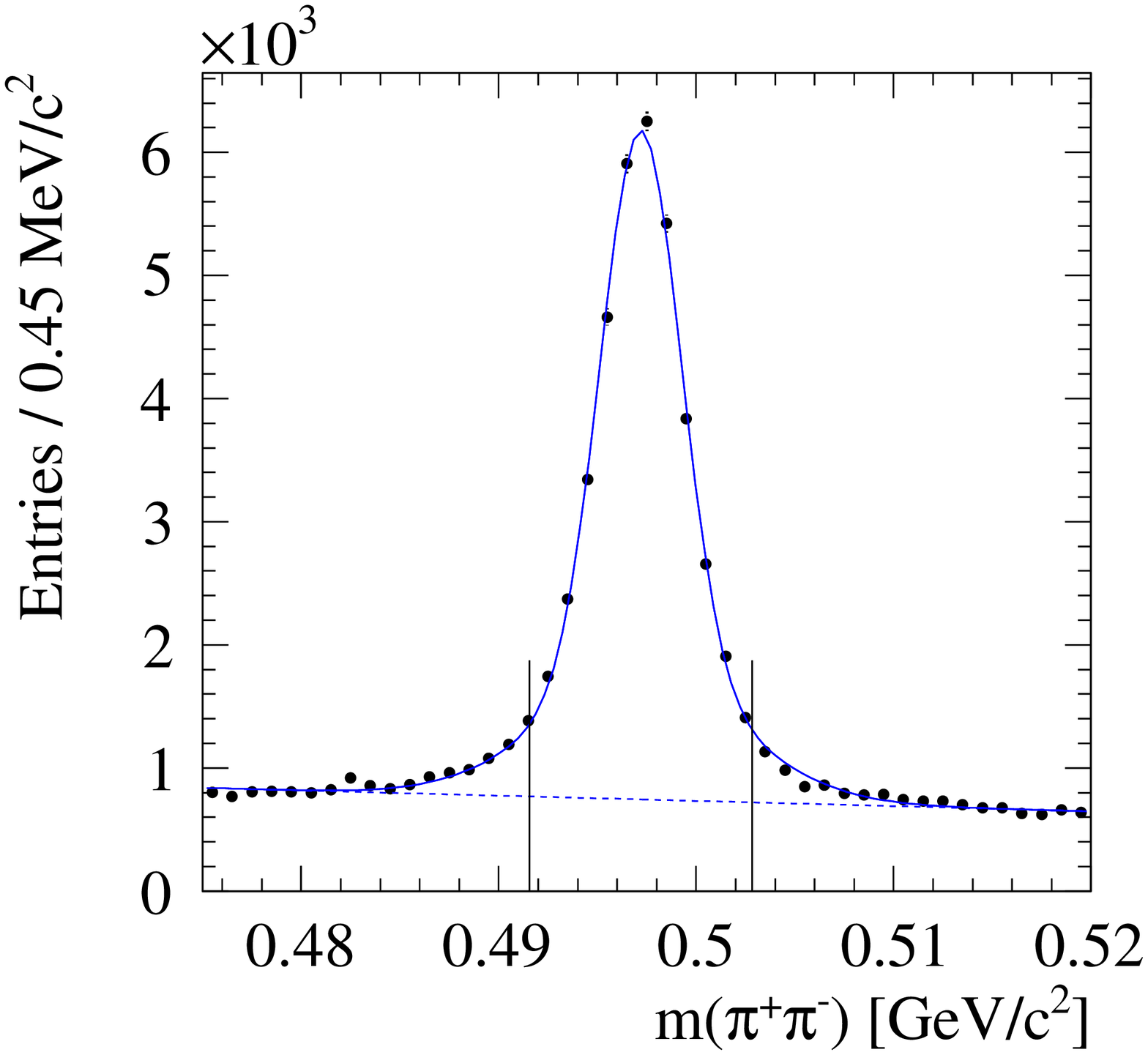}\\
\includegraphics[width=0.3\textwidth]{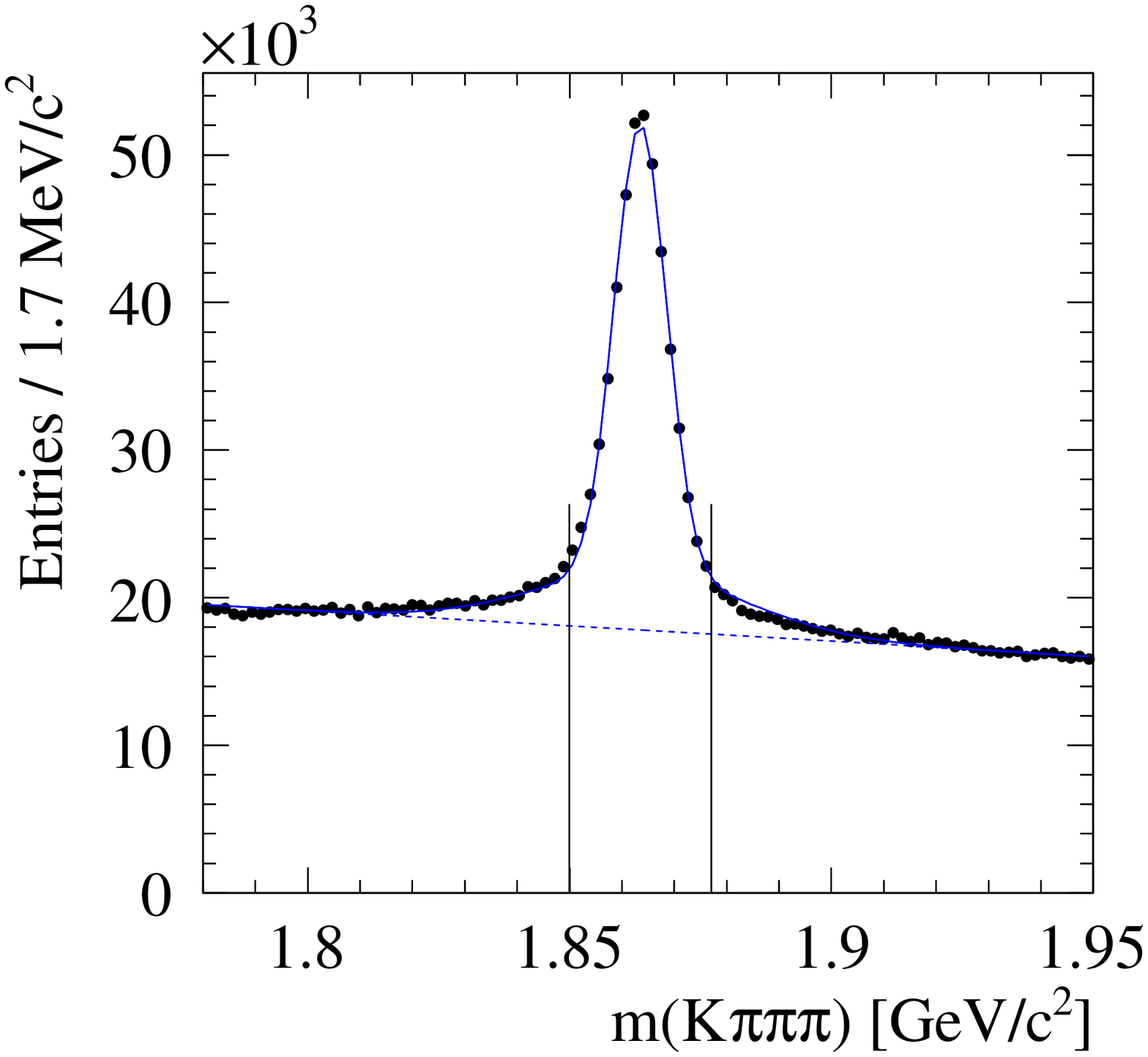}
\includegraphics[width=0.3\textwidth]{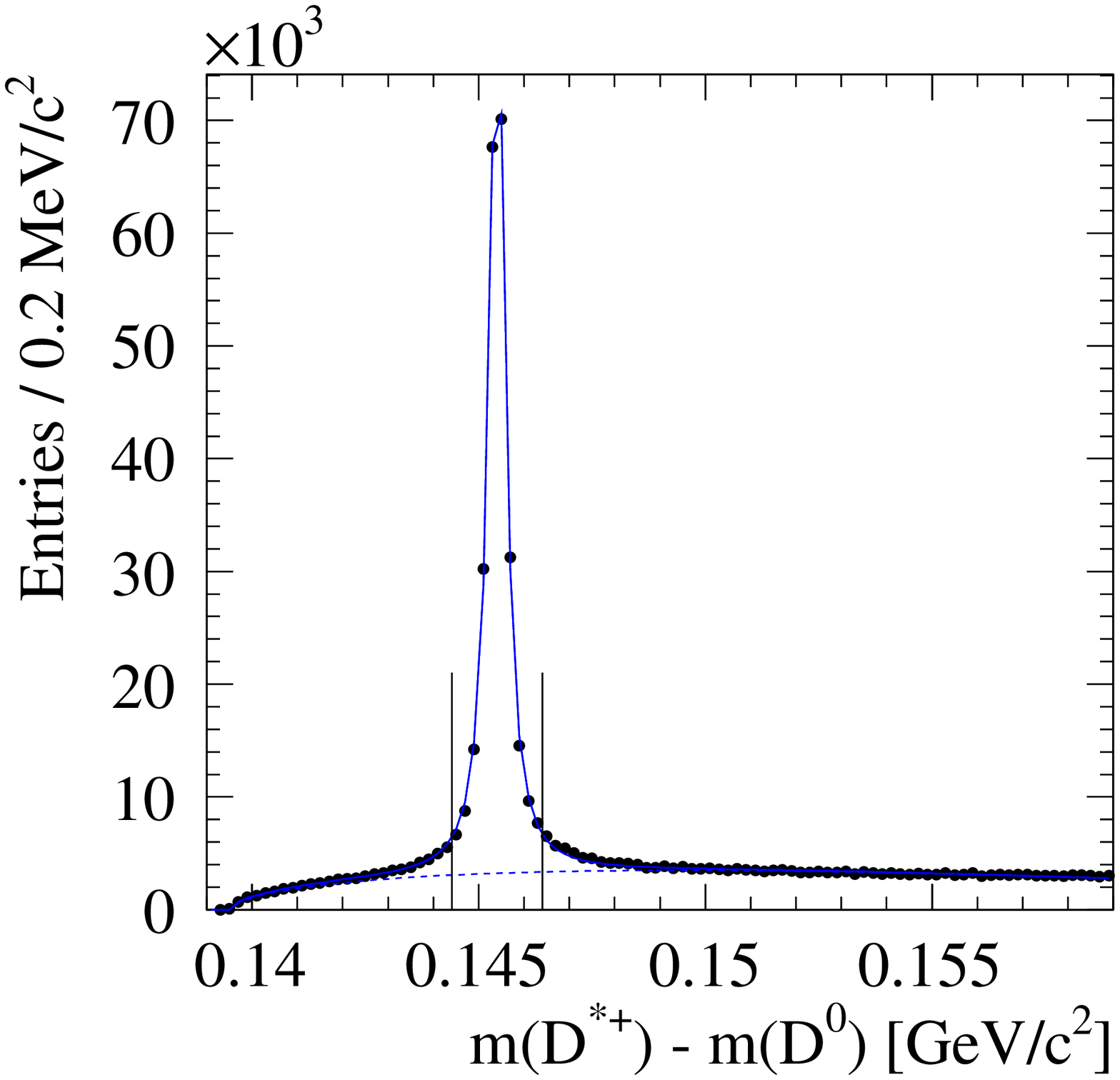}
\includegraphics[width=0.3\textwidth]{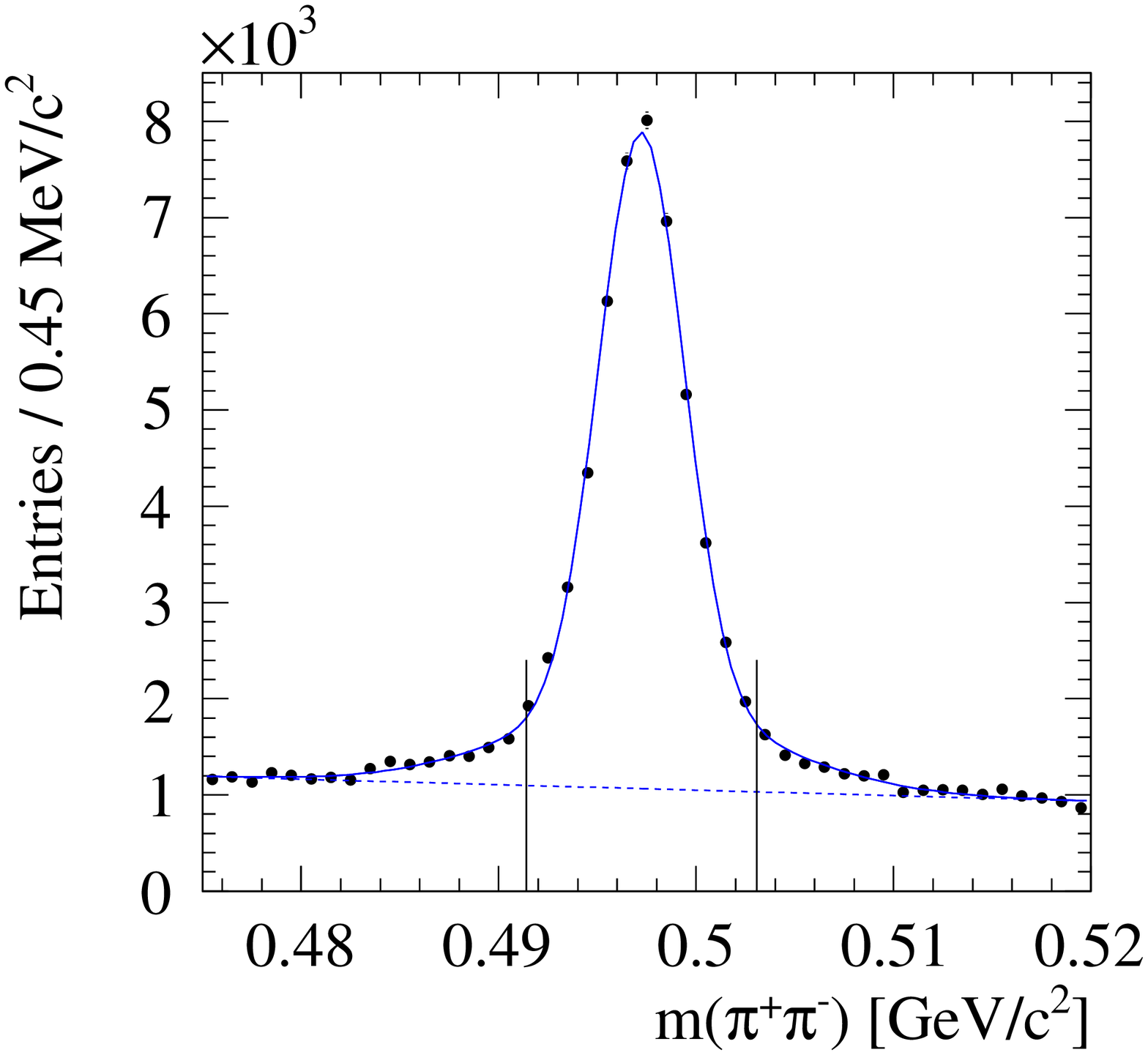}
\begin{picture}(0,0)
\put(-12.,9.3){(a)}
\put(-12.,4.2){(d)}
\put(-6.5,9.3){(b)}
\put(-6.5,4.2){(e)}
\put(-1.,9.3){(c)}
\put(-1.,4.2){(f)}
\end{picture}
\caption{(a,d) \Dz candidate invariant mass distributions; (b,e) Difference between the \Dstarp and \Dz candidate invariant masses. (c,f) \KS candidate invariant mass distributions; (a-c) $K4\pi$ mode; (d-f) $K6\pi$ mode. Signal regions are indicated by the vertical lines. The signal and background line shapes fitted to the mass distributions are described in the text.}
\label{fig:recores}
\end{figure*}

\begin{figure}
\includegraphics[width=0.3\textwidth]{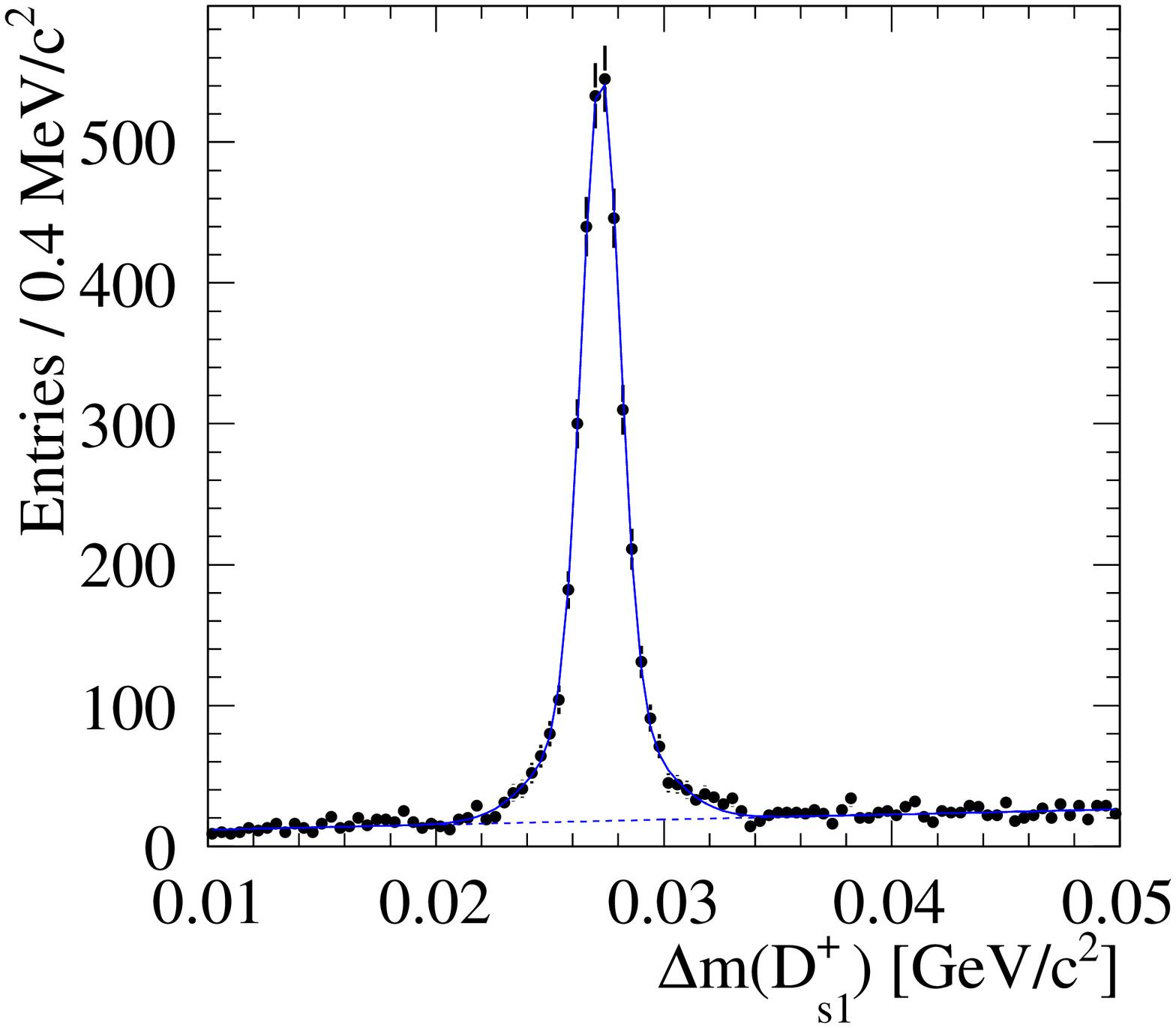}
\includegraphics[width=0.3\textwidth]{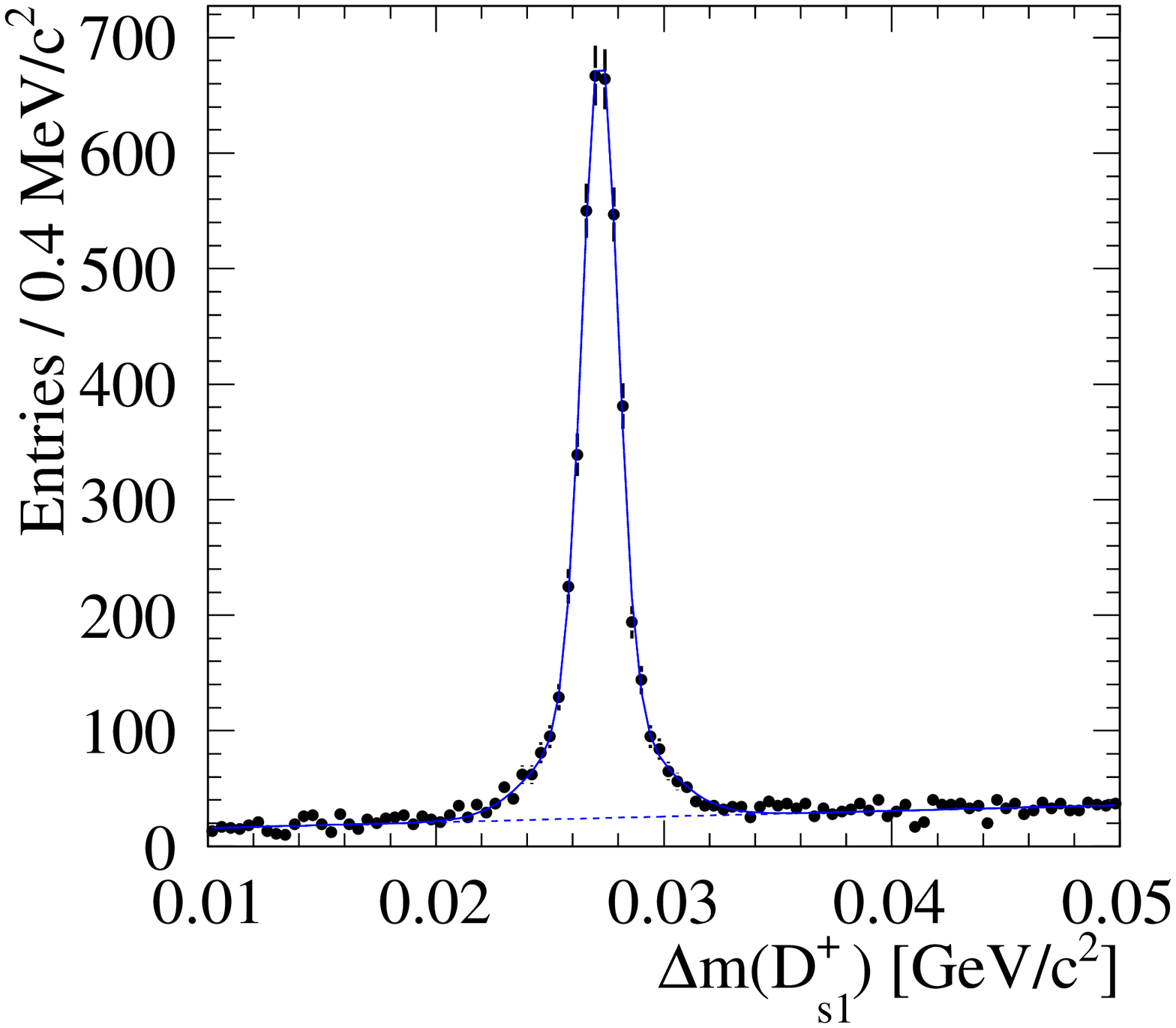}
\begin{picture}(0,0)
\put(-1.,9.3){(a)}
\put(-1.,4.2){(b)}
\end{picture}
\caption{$\dmds = m(\Dsop) - m(\Dstarp) - m(\KS)$ invariant mass distributions in data after applying all selection criteria for the (a) $K4\pi$ and (b) $K6\pi$ mode. A Double-Gaussian signal function and a linear background function are used to describe the data in a preliminary fit.}
\label{fig:ds1mass}
\end{figure}

The \Dsop is reconstructed via its decay mode $\Dstarp\KS$, with $\KS \to \pip\pim$ and $\Dstarp \to \Dz\pip$. The \Dz is reconstructed through two decay modes, $\Km\pip$ and $\Km\pip\pip\pim$, which will be labeled $K4\pi$ and $K6\pi$, respectively, in the following. To improve the mass resolution, the mass difference $\dmds = m(\Dsop) - m(\Dstarp) - m(\KS)$ is examined.\\
\indent Events are selected by requiring at least five charged tracks, at least one of which is identified as a charged kaon. Also, at least one neutral kaon candidate is required. Each track must approach the nominal \epem interaction point to within $1.5 \cm$ in the transverse direction, and to within $10 \cm$ in the longitudinal (beam) direction. Kaon candidates are identified using the normalized kaon, pion and proton likelihood values ($L_{K}$, $L_{\pi}$ and $L_{p}$) obtained from the particle identification system, by requiring $L_{K}/(L_{K}+L_{\pi}) > 0.50$ and $L_{K}/(L_{K}+L_{p}) > 0.018$. Furthermore, the track must be inconsistent with the electron hypothesis or have a momentum less than $0.4\gevc$. Tracks that fulfill $L_{K}/(L_{K}+L_{\pi}) < 0.98$ and $L_{p}/(L_{p}+L_{\pi}) < 0.98$ are selected as pions. \\
\indent Candidates for the \Dz decay are formed by selecting all $\Km\pip$ pairs ($\Km\pip\pip\pim$ combinations in the second mode) that have an invariant mass within $\pm 100~\mevcc$ of the nominal mass~\cite{Pd10}. Candidates for the \Dstarp decay are formed by adding a \pip to the \Dz, such that the mass difference between \Dstarp and \Dz is less than $170\mevcc$. A \KS candidate consists of a \pip\pim pair with invariant mass within $\pm 25\mevcc$ of the nominal mass~\cite{Pd10}. A kinematic fit is applied to the complete decay chain, constraining the \Dsop candidate vertex to be consistent with the \epem interaction region. Mass constraints are not applied to intermediate states. Those \Dsop candidates with a $\chi^{2}$ fit probability greater than $0.1 \%$ are retained. To suppress combinatorial background and events from $B$-decays, we require the momentum $\pstar$ of the \Dsop in the \FourS center-of-mass (CM) frame to exceed $2.7 \gevc$. \\
\indent The $K\pi$ and $K\pi\pi\pi$ mass spectra for accepted \Dz candidates, shown in Figs.~\ref{fig:recores}(a) and~\ref{fig:recores}(d), are fitted with a signal function consisting of a sum of two Gaussians with a common mean value, and a linear background function. The width of the signal regions for $\Dz$, $\Dstarp$ and $\KS$ candidates is defined as twice the full width at half maximum (FWHM) of the signal line shapes. A signal window of $\pm 18~(14)\mevcc$ for the $K4\pi$~($K6\pi$) mode around the mean mass of $1863.5~(1863.5)\mevcc$ obtained from the fit is used to select \Dz candidates. For these candidates, the $\Dz\pip - \Dz$ mass difference distributions are shown in Figs.~\ref{fig:recores}(b) and~\ref{fig:recores}(e). These are fitted with the sum of a relativistic Breit-Wigner signal function and a background function consisting of a polynomial times an exponential function. A \Dstarp signal region of $\pm 1\mevcc$ around the fitted mean value of $145.4\mevcc$ is chosen for both decay modes. To further reduce the background, the angle between the flight direction of the \KS candidate and the line connecting the \epem interaction point and the \KS decay vertex is required to be less than $0.15~\mathrm{radians}$. For candidates passing these selection criteria, the \KS candidate invariant mass distributions (Figs.~\ref{fig:recores}(c) and~\ref{fig:recores}(f)) are fitted with the sum of a signal function, consisting of the sum of two Gaussians, and a linear background function. A signal window of $\pm 6\mevcc$ around the fitted mean mass of $497.2\mevcc$ is selected for both decay modes.   \\
\indent In the case of an event with multiple candidates, the candidate with the best fit probability is chosen. The \dmds candidate spectra after all selection criteria are shown in Figs.~\ref{fig:ds1mass}(a) and~\ref{fig:ds1mass}(b). The fits to these spectra use a Double-Gaussian signal function and a linear background function. Note that for this preliminary fit the intrinsic width and the resolution are not taken into account. The FWHM values obtained are $(2.2 \pm 0.1)~\mev$ and $(2.0\pm0.1)~\mev$, respectively, with corresponding signal yields of about $3500$ and $4000$ entries.

\section{Monte Carlo simulation and Comparison with Data}
\label{sec:mc}
\begin{figure}
\includegraphics[width=0.3\textwidth]{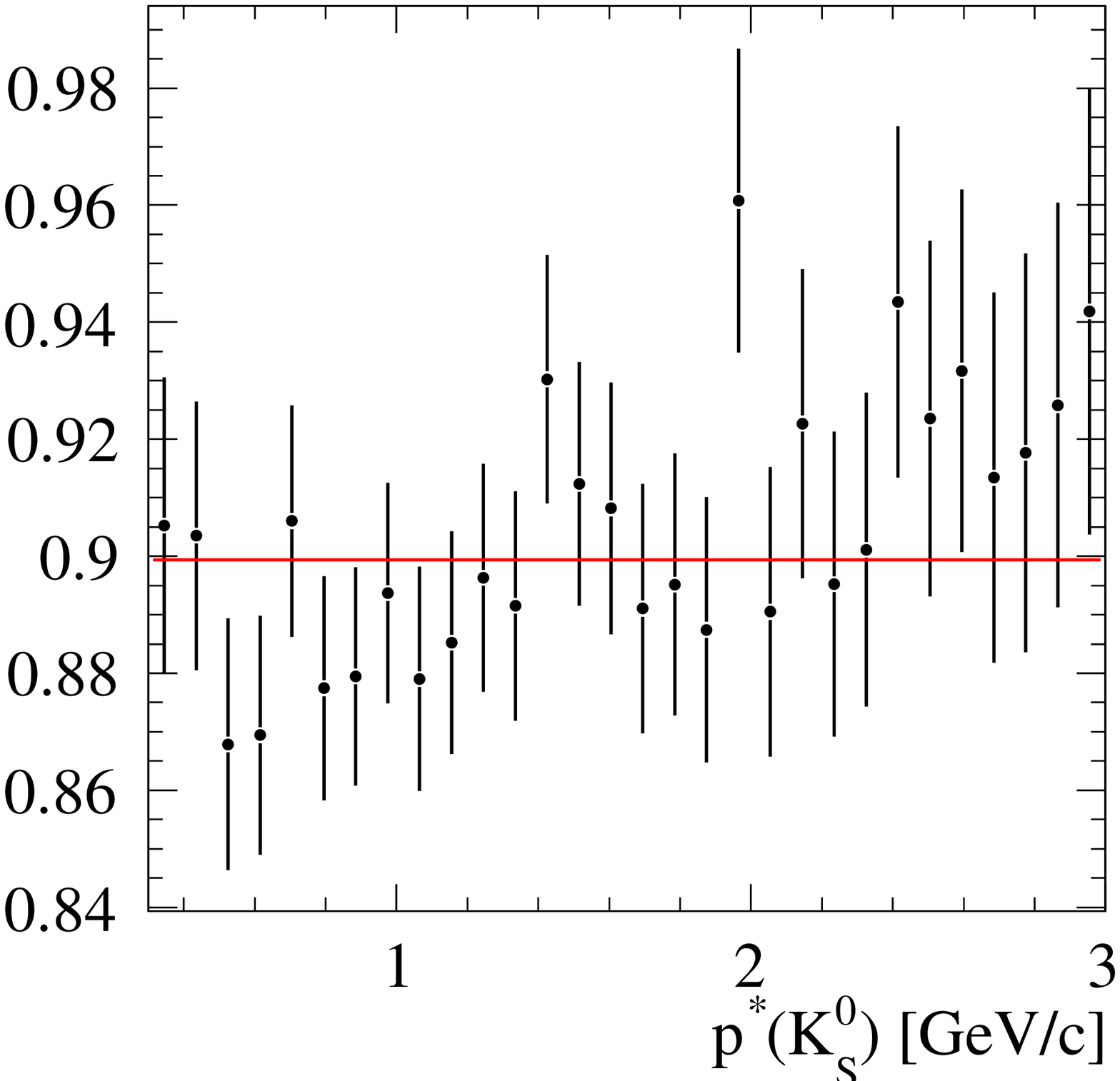}
\includegraphics[width=0.3\textwidth]{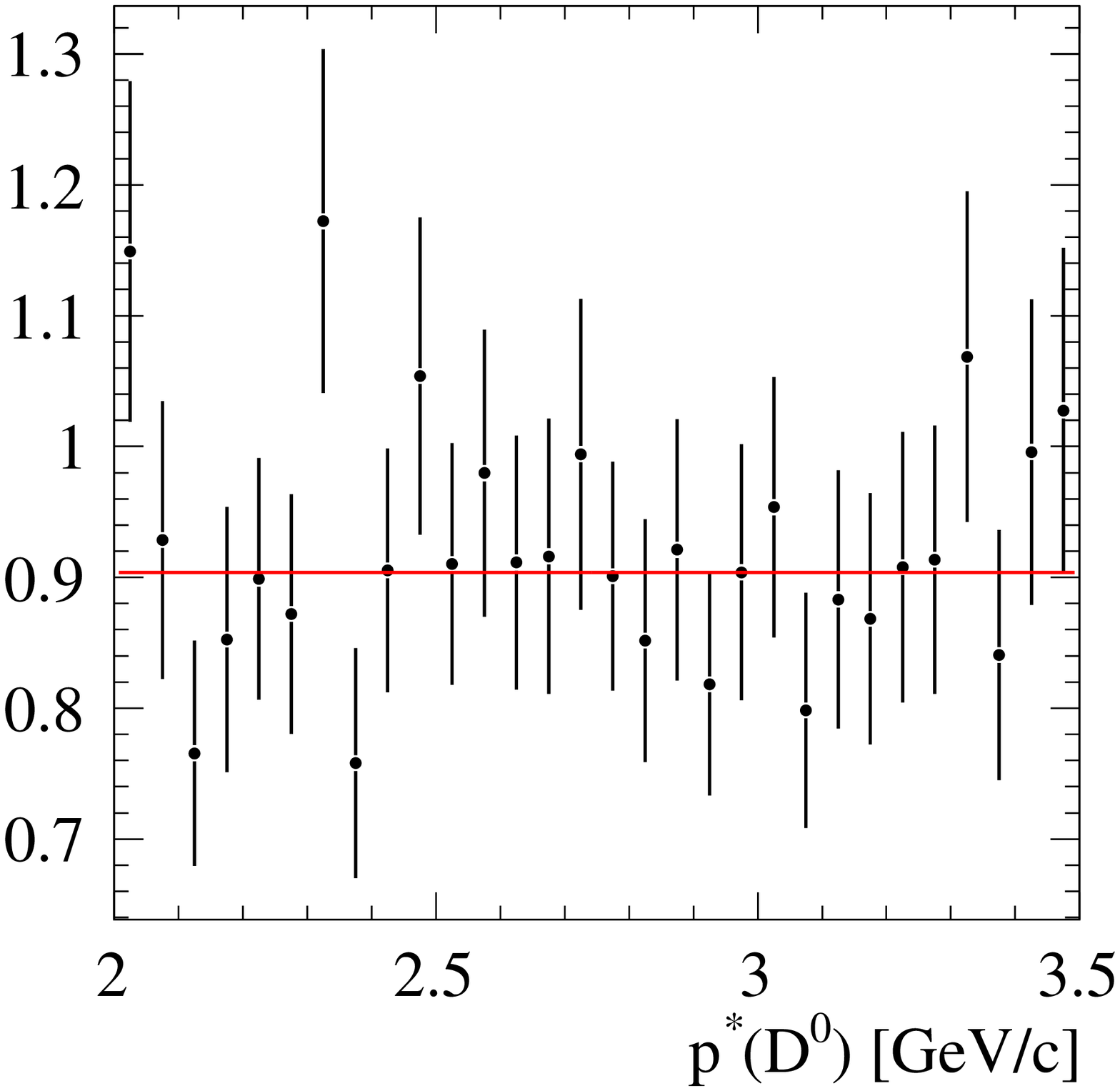}
\begin{picture}(0,0)
\put(-4.5,9.3){(a)}
\put(-4.5,4.2){(b)}
\put(-5.6,8.){\begin{rotate}{90} {\large $\frac{\mathrm{FWHM(MC)}}{\mathrm{FWHM(Data)}}$} \end{rotate}}
\put(-5.6,2.8){\begin{rotate}{90} {\large $\frac{\mathrm{FWHM(MC)}}{\mathrm{FWHM(Data)}}$} \end{rotate}}
\end{picture}
\caption{$p^{\ast}$-dependence of the ratio between the FWHM of the resolution functions from \ccbar-MC and data. (a) $\KS \to \pip\pim$. (b) $\Dz \to \Km\pip\pip\pim$. The solid line shows the fitted mean ratio with a value of $~0.9$.}
\label{fig:resrat}
\end{figure}

\begin{figure}
\includegraphics[width=0.3\textwidth]{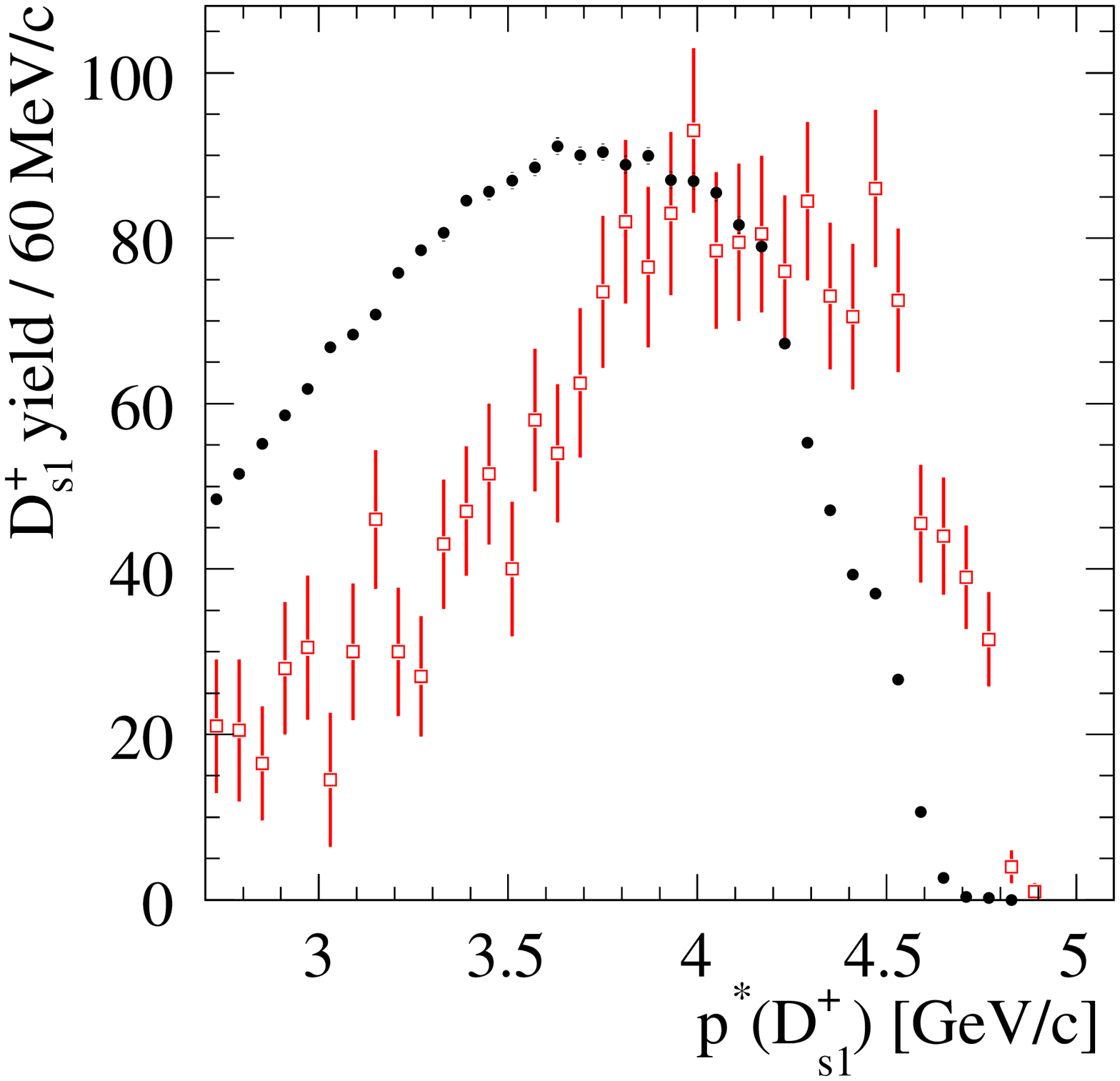}
\includegraphics[width=0.3\textwidth]{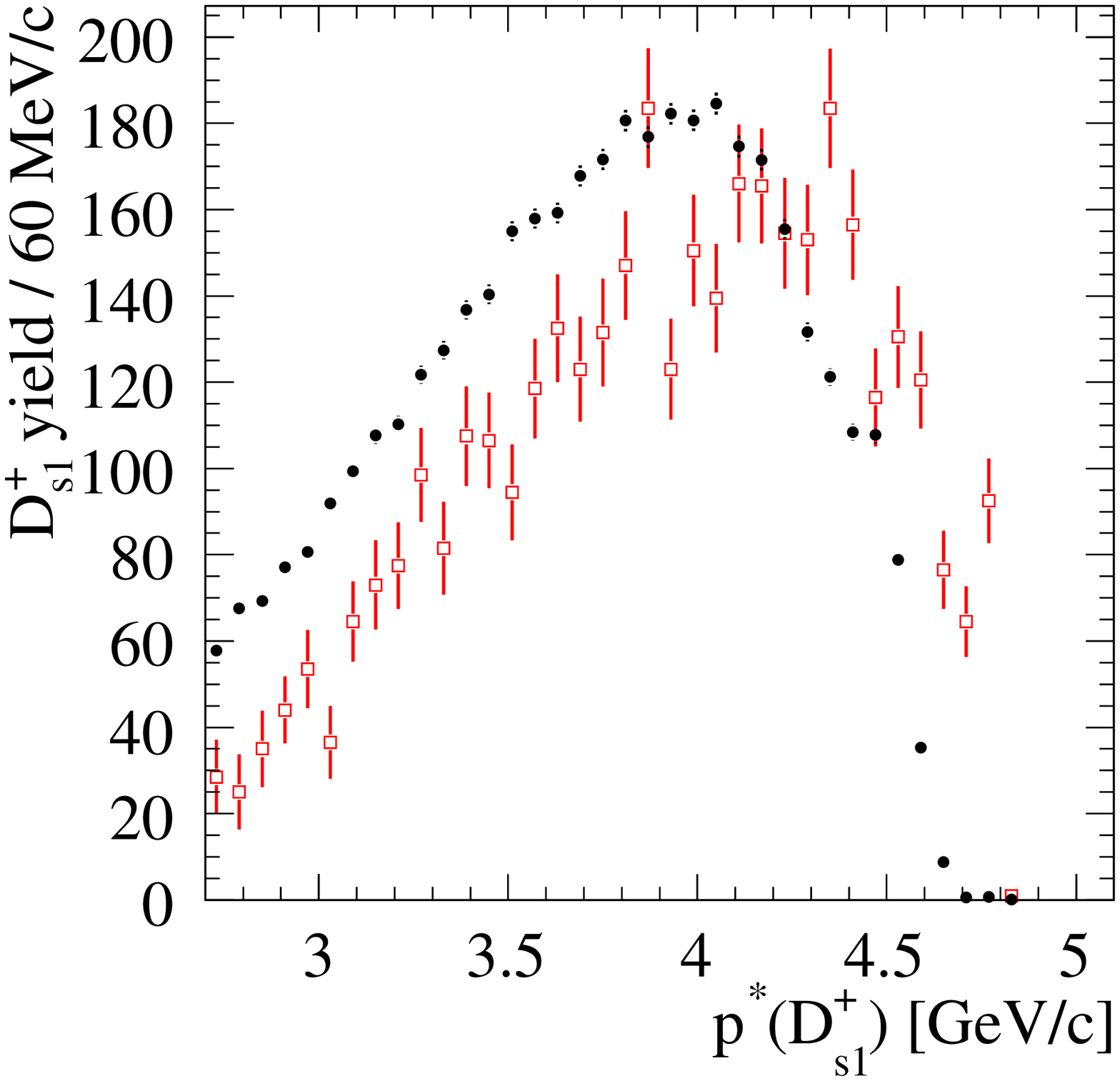}
\begin{picture}(0,0)
\put(-4.3,9.4){(a)}
\put(-4.3,4.3){(b)}
\end{picture}
\caption{\pstar-dependence of the \Dsop signal yield for data (open squares) and reconstructed MC (solid points) for the (a) $K4\pi$ and (b) $K6\pi$ decay modes.}
\label{fig:pstardiff}
\end{figure}

Monte Carlo events are generated for $\Dsop \to \Dstarp\KS, \Dstarp\to\Dz\pip$, with $\Dz \to \Km\pip$ and $\Dz\to\Km\pip\pip\pim$, by {\tt EvtGen}~\cite{La01}. The detector response is simulated using the {\tt GEANT4}~\cite{Ag03} package. For each \Dz decay mode, and for each of the corresponding \Dsom decays, $776000$ events are generated. The \Dsop line shape is generated using a non-relativistic Breit-Wigner function having central value $m(\Dsop)_{gen} = 2535.35\mevcc$ and intrinsic width $\Gamma(\Dsop)_{gen} = 1\mev$ (this sample is labeled $\Gamma_{1}$ in the following). The range of generated \Dsop masses is restricted to values between $m(\Dsop)_{gen} - 10\mevcc$ and $m(\Dsop)_{gen} + 15\mevcc$. The masses of the daughter particles are taken from Ref.~\cite{Pd10}. \\
\indent In order to test the mass resolution model, a second set of MC samples with $381000$ events for each \Dz decay mode is generated using a Breit-Wigner width of $\Gamma(\Dsop)_{gen} = 2\mev$~($\Gamma_{2}$ sample). In addition to these signal MC samples, separate $\Dz$ and $\KS$ samples are created from data and generic \ccbar MC simulations without requiring a \Dstarp or \Dsop. They are used mainly for resolution studies.\\
\indent The MC and data are in good agreement for the transverse momentum distributions of pions, kaons, $D$ and $\Dstar$ mesons, and for the number of SVT coordinates of pions and kaons. The agreement is worse for the number of DCH coordinates, where the data show systematically fewer coordinates than the MC, giving rise to a resolution that is about $10\%$ smaller in the MC than in data. This is illustrated in Fig.~\ref{fig:resrat}, which shows the $p^{\ast}(\KS)$ and $p^{\ast}(\Dz)$ dependence of the ratio between the FWHM of the resolution functions in \ccbar MC and data, where $p^{\ast}$ is the momentum in the CM frame. This effect will be discussed further in Sec.~\ref{sec:syst}. There is also disagreement between the number of \Dsop signal entries in MC and data as a function of \pstar (Fig.~\ref{fig:pstardiff}). This effect will be addressed in Secs.~\ref{sec:res} and~\ref{sec:syst}.

\section{Resolution Model}
\label{sec:res}
\begin{figure}
\includegraphics[width=0.3\textwidth]{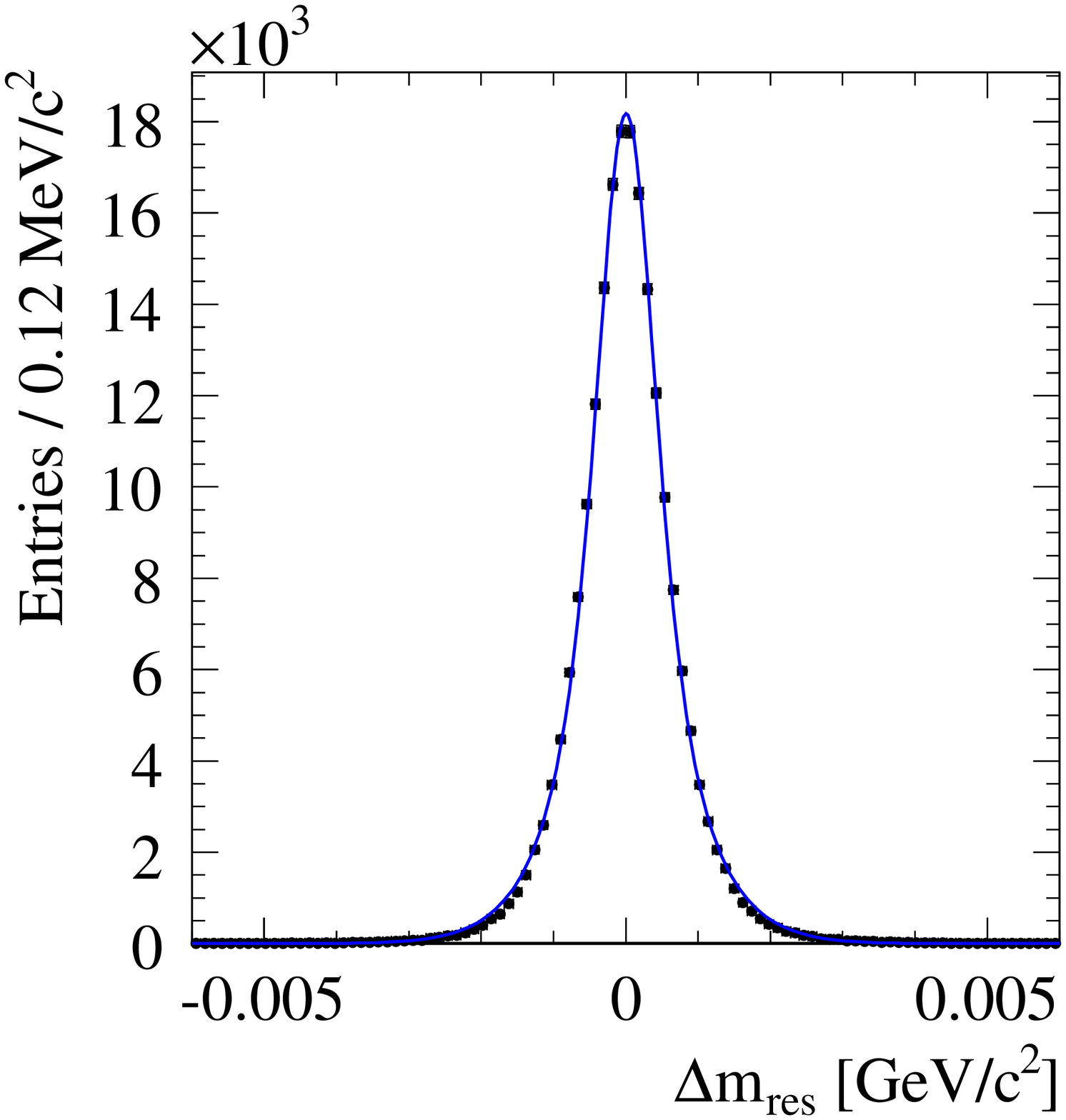}
\includegraphics[width=0.3\textwidth]{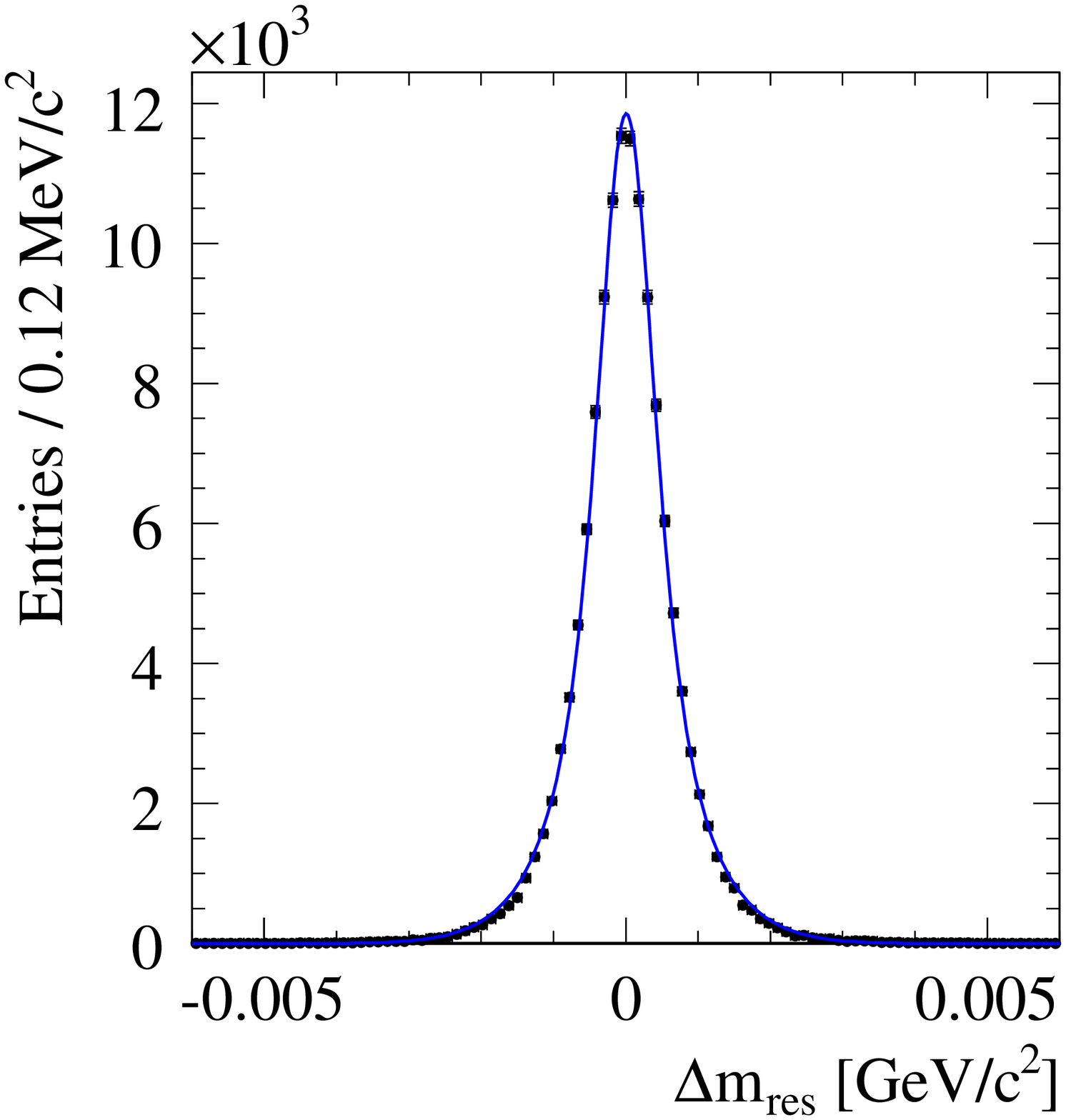}
\begin{picture}(0,0)
\put(-4.1,9.5){(a)}
\put(-4.1,4.2){(b)}
\end{picture}
\caption{Fit of the resolution function (Eq.~(\ref{eq:mulg})) to $\Delta m_{res}$ with the $r$ and $\sigma_{0}$ parameters free to vary for the (a) $K4\pi$ and (b) $K6\pi$ decay modes.}
\label{fig:res_rfit}
\end{figure}

\begin{figure}
\includegraphics[width=0.33\textwidth]{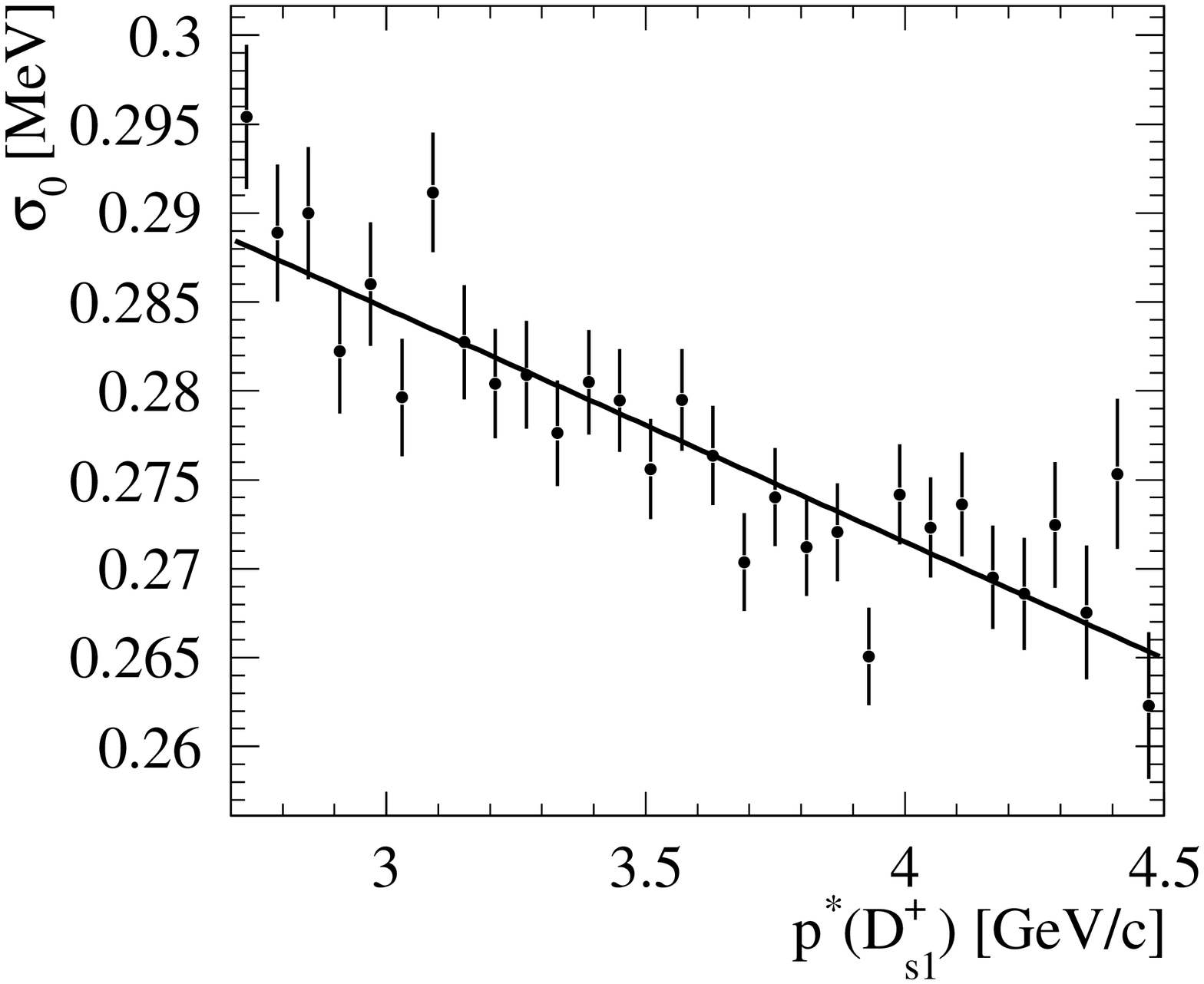}
\includegraphics[width=0.33\textwidth]{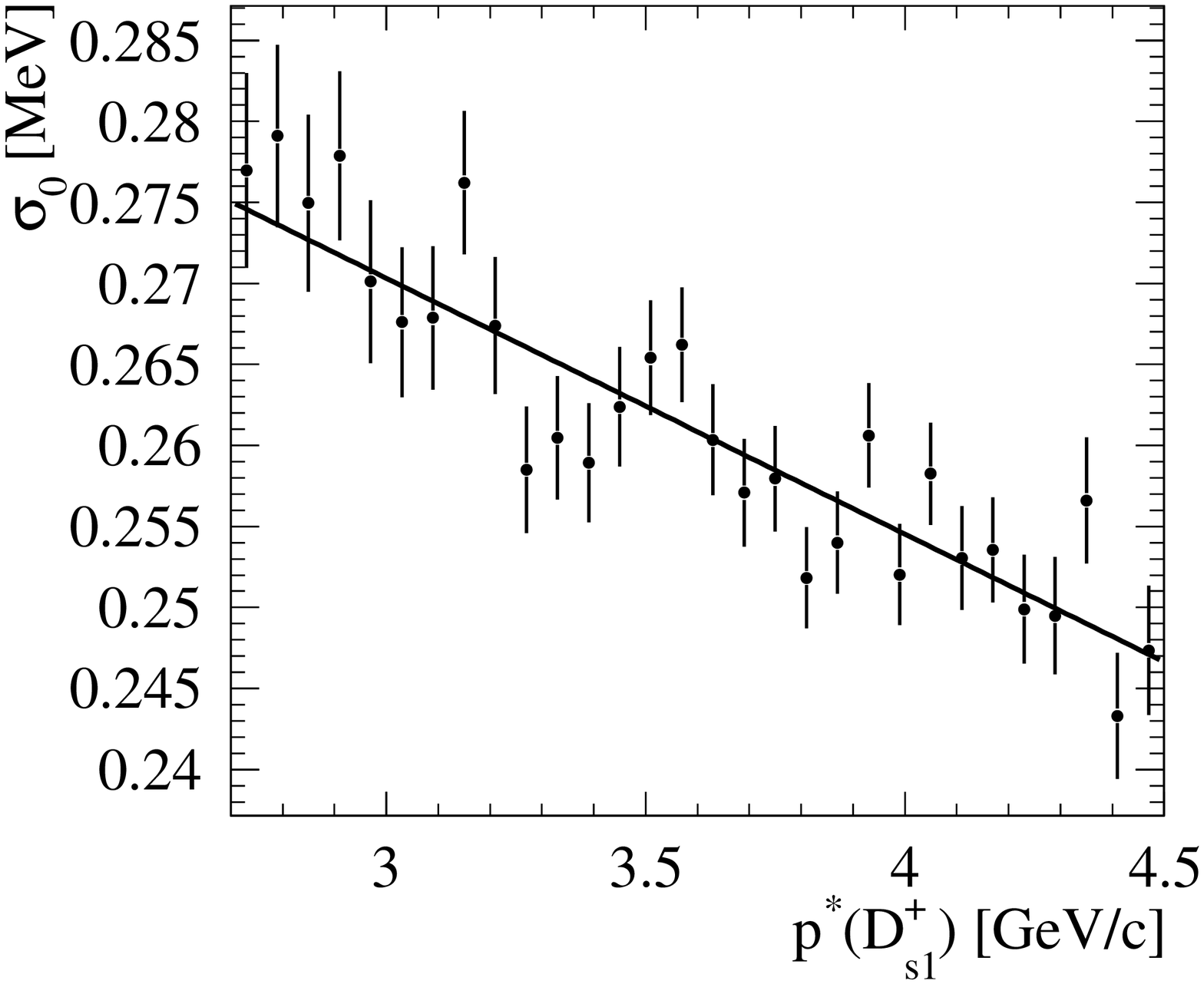}
\begin{picture}(0,0)
\put(-1.,9.3){(a)}
\put(-1.,4.2){(b)}
\end{picture}
\caption{\pstar dependence of the resolution function parameter $\sigma_{0}$, represented by a linear parametrization~($r$ fixed) for the (a) $K4\pi$ and (b) $K6\pi$ decay modes.}
\label{fig:res_sigma}
\end{figure}

\begin{figure}
\includegraphics[width=0.3\textwidth]{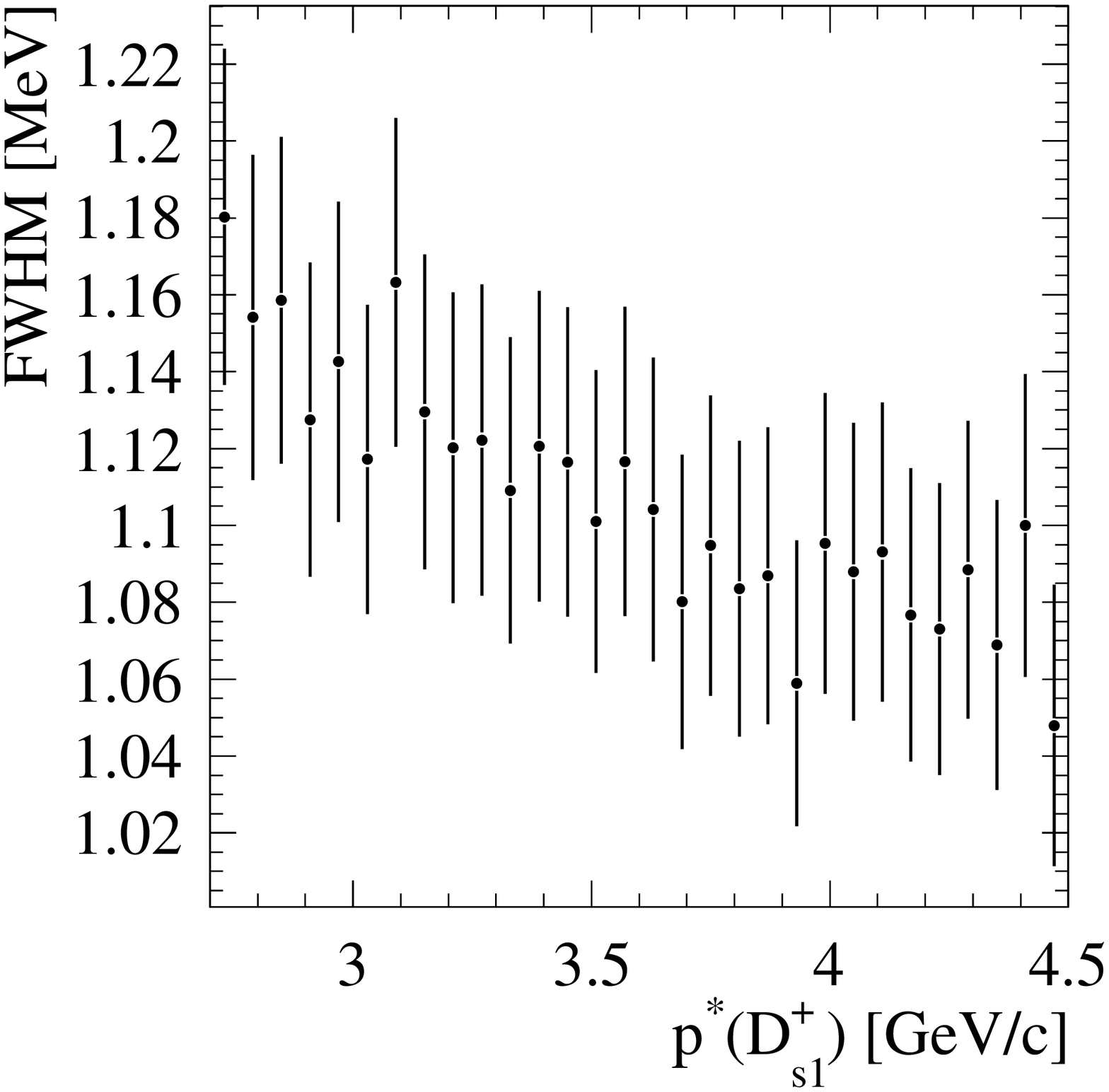}
\includegraphics[width=0.3\textwidth]{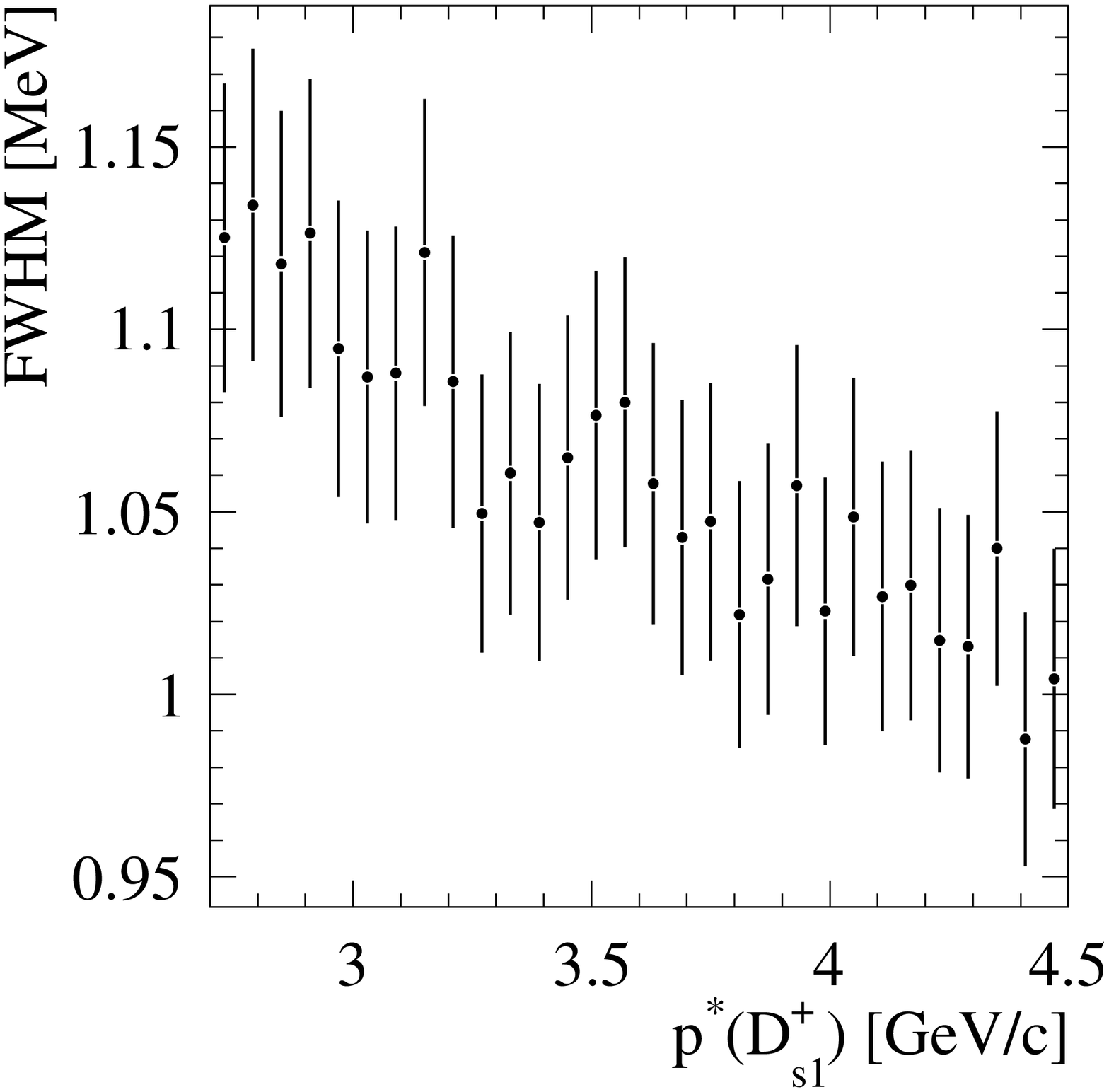}
\begin{picture}(0,0)
\put(-1.,10.3){(a)}
\put(-1.,4.5){(b)}
\end{picture}
\caption{\pstar dependence of the FWHM of the resolution function~($r$ fixed) for the (a) $K4\pi$ and (b) $K6\pi$ decay modes.}
\label{fig:res_fwhm}
\end{figure}

\begin{figure}
\includegraphics[width=0.3\textwidth]{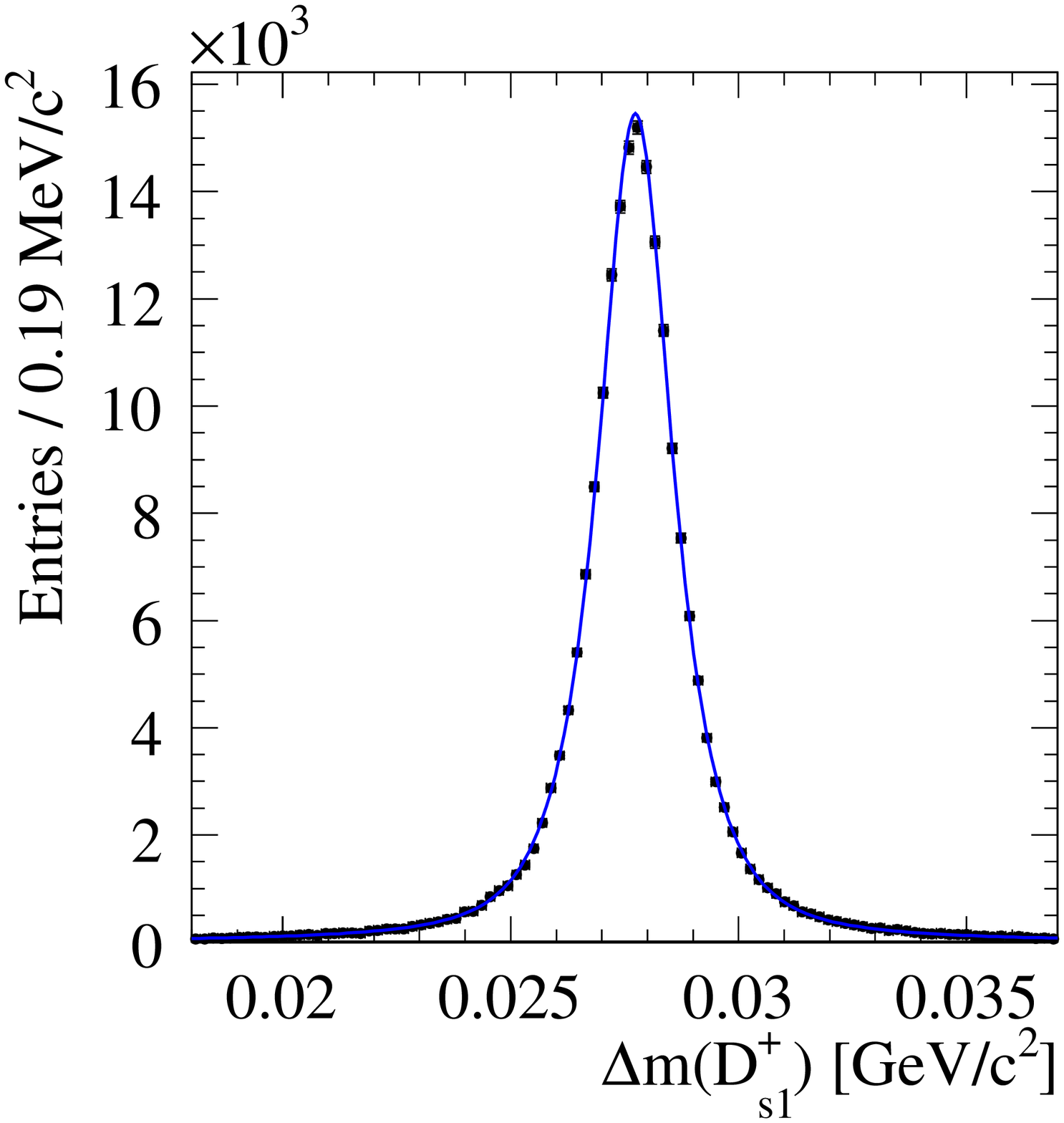}
\includegraphics[width=0.3\textwidth]{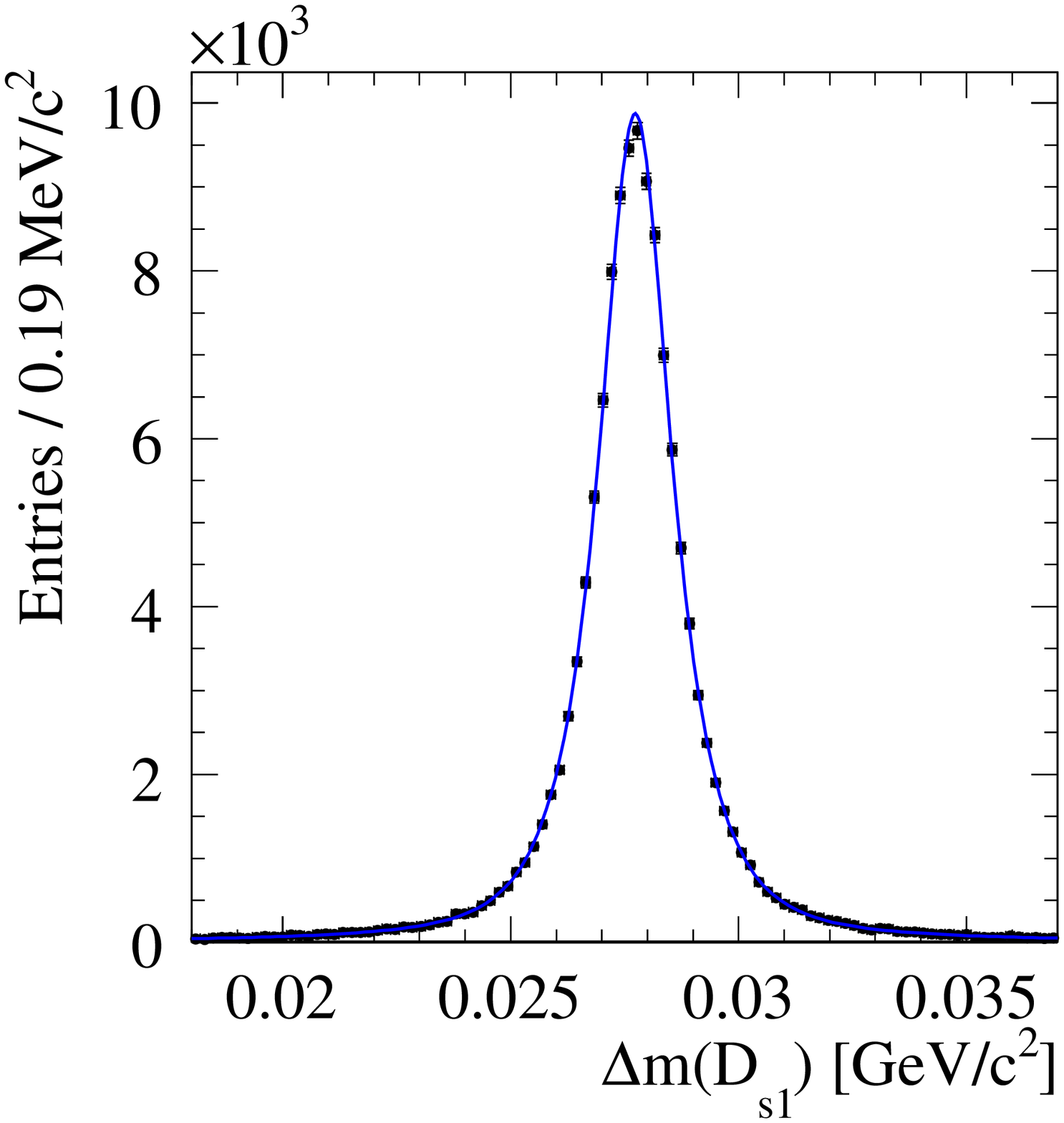}
\begin{picture}(0,0)
\put(-1.,9.3){(a)}
\put(-1.,4.2){(b)}
\end{picture}
\caption{Fit of a non-relativistic Breit-Wigner convolved with the resolution function to the \Dsop candidate mass difference spectra in the $\Gamma_{1}$ MC sample for the (a) $K4\pi$ and (b) $K6\pi$ decay modes.}
\label{fig:MCtestfit}
\end{figure}

The resolution model is derived from the \Dsop signal MC by studying the difference $\Delta m_{res}$ between the reconstructed and generated \Dsop mass values. The Multi-Gaussian ansatz \begin{equation}
 G(\Delta m_{res})=\int_{\sigma_0}^{r\sigma_0} \frac{1}{r\sigma^2}e^{-\frac{(\Delta m_{res}-\Delta m_{res0})^2}{2\sigma^2}}\mathrm{d}\sigma
\label{eq:mulg}
\end{equation}
is found to accurately model the mass resolution spectra. This represents a superposition of Gaussian distributions with the same mean value $\Delta m_{res0}$ but variable width $\sigma$, starting from minimum width $\sigma_{0}$ and increasing to maximum width $r\sigma_{0}$. The FWHM of the distribution is numerically calculable once $\sigma_{0}$ and $r$ are known. The mass resolution for the different particles depends on the CM momentum \pstar. Therefore, the parameter $\sigma_{0}$ of Eq.~(\ref{eq:mulg}) is obtained as a function of \pstar.\\
\indent  Figures~\ref{fig:res_rfit}(a) and~\ref{fig:res_rfit}(b) show $\Delta m_{res}$ distributions for the full \pstar range. From these plots the value of the parameter $r$ is determined to be $4.78 \pm 0.04$ and $5.20 \pm 0.05$ for the $K4\pi$ and $K6\pi$ modes, respectively. Events are divided into 30 \pstar intervals from $2.7\gevc$ to $4.7\gevc$ and the fit repeated for each interval, resulting in \pstar-dependent $\sigma_{0}$ values (Figs.~\ref{fig:res_sigma}(a) and~\ref{fig:res_sigma}(b)). The corresponding \pstar-dependent FWHM of the resolution functions is shown in Figs.~\ref{fig:res_fwhm}(a) and~\ref{fig:res_fwhm}(b). \\
\indent In order to validate this resolution model, the \pstar-dependent resolution function with the corresponding parameters $\sigma_0$ and $r$ is convolved with a non-relativistic Breit-Wigner function and fitted to the \dmds signal MC distribution (MC sample $\Gamma_{1}$). The results are shown in Figs.~\ref{fig:MCtestfit}(a) and~\ref{fig:MCtestfit}(b), and the reconstructed values for mean \dmdsz and width \gds are listed in Table~\ref{tab:testfit}. The corresponding generated values for both decay modes are $\dmds_{gen} = 27.744\mevcc$ for the mean and $\gds_{gen}=1.000\mev$ for the width. The small deviations between generated and reconstructed values are discussed in Sec.~\ref{sec:syst}.

\begin{table}
\caption{Reconstructed values for \dmdsz and \gds (fit to MC sample $\Gamma_{1}$). The resolution model used is derived from MC sample $\Gamma_{1}$.}
\begin{ruledtabular}
  \begin{tabular}{lcc}
   & $\dmdsz / \mevcc$ & $\gds / \mev$ \\ \hline
   $(K4\pi)$ & $27.737 \pm 0.003$ & $1.001 \pm 0.005$ \\ 
   $(K6\pi)$ & $27.734 \pm 0.003$ & $0.991 \pm 0.006$\\ 	
     \end{tabular}
\label{tab:testfit}
\end{ruledtabular}
\end{table}

\section{Fit to the {\boldmath $\Dstar\KS$} mass spectrum}
\label{sec:datafit}

\begin{figure}
\includegraphics[width=0.38\textwidth]{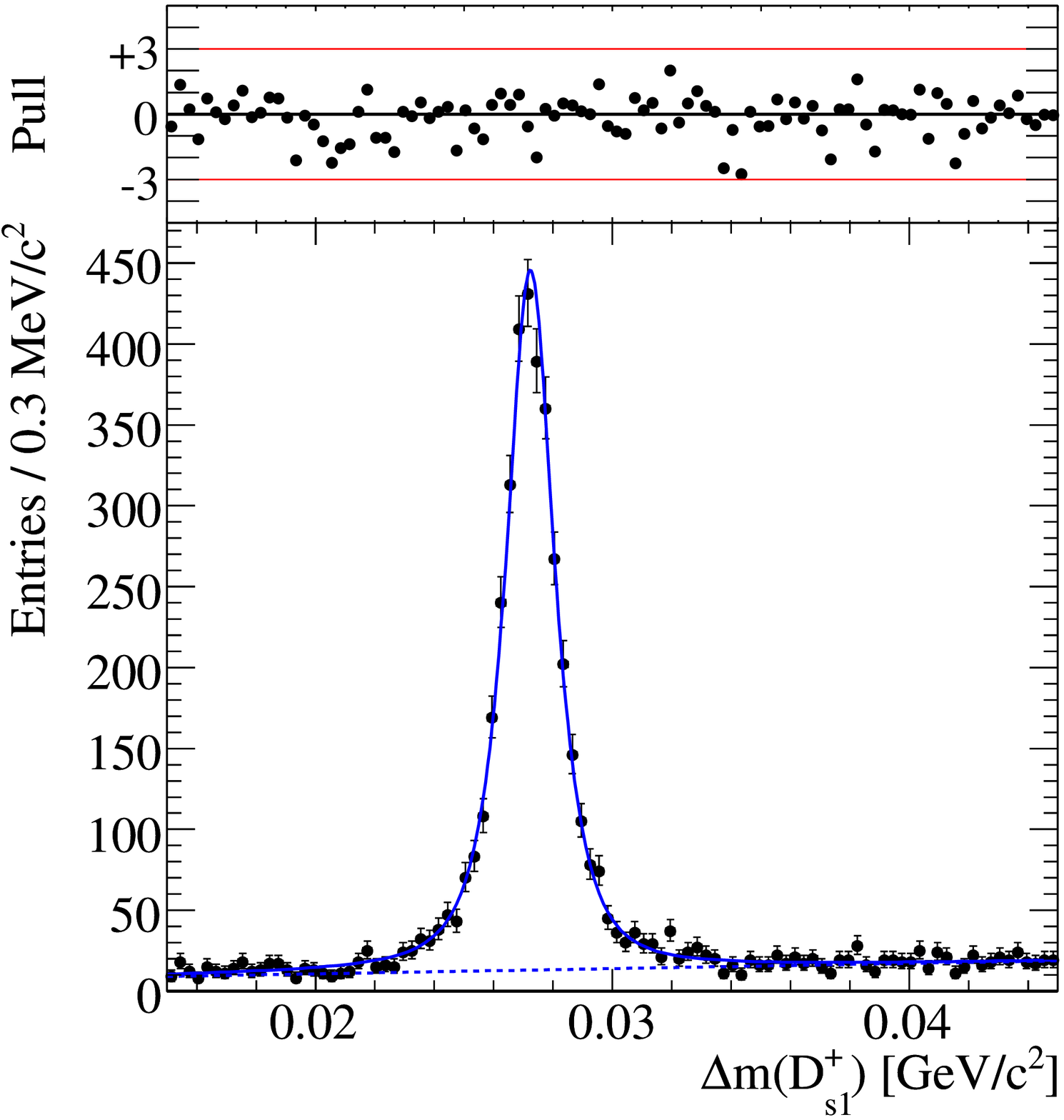}
\includegraphics[width=0.38\textwidth]{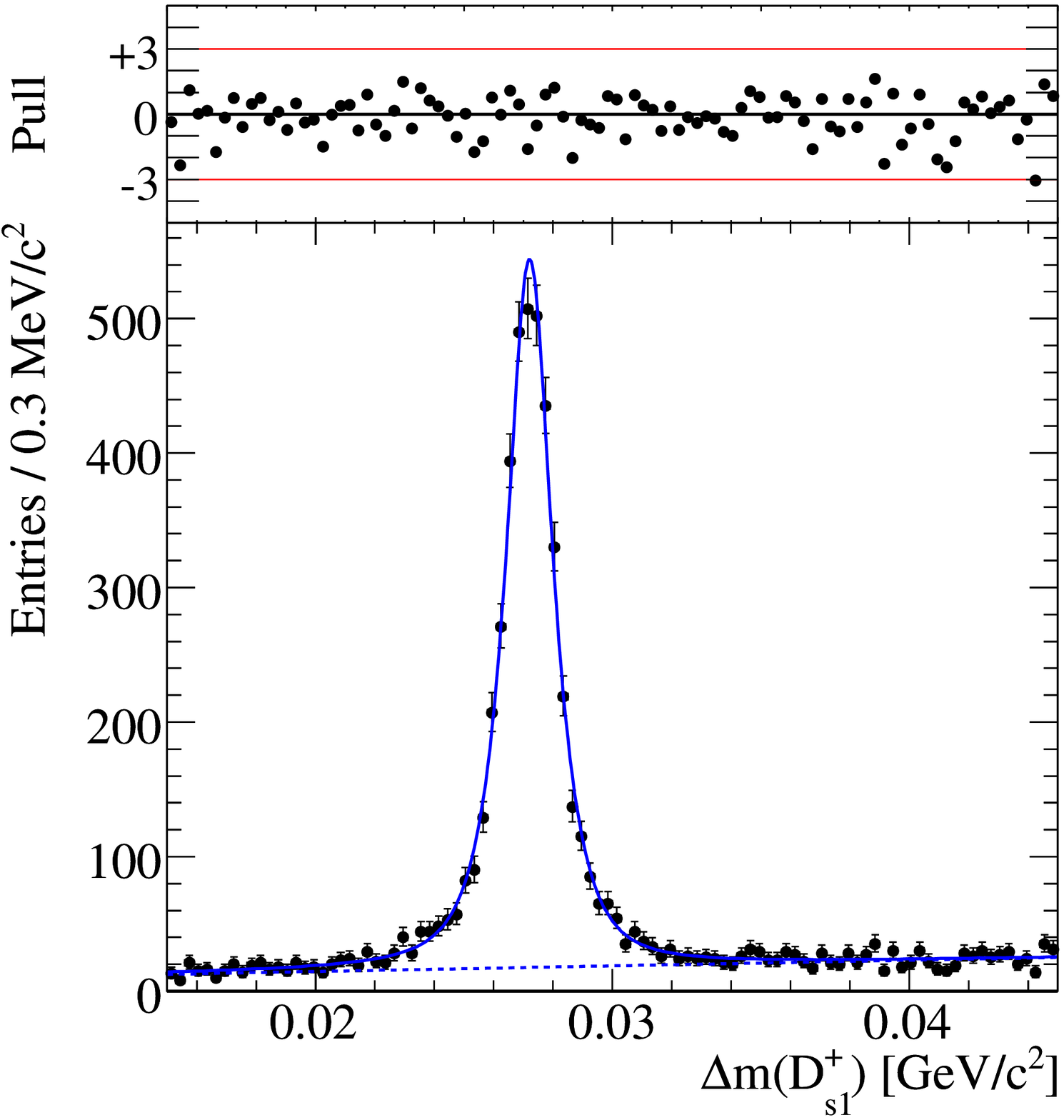}
\begin{picture}(0,0)
\put(-1.5,11.8){(a)}
\put(-1.5,4.8){(b)}
\end{picture}
\caption{Fit of a relativistic Breit-Wigner convolved with the resolution function to the \Dsop candidate mass difference spectra in data, for the (a) $K4\pi$ and (b) $K6\pi$ modes. The dotted line indicates the background line shape. The upper parts of the figures show the normalized fit residuals.}
\label{fig:datafit}
\end{figure}

For the final fit to the $\Dstar\KS$ mass spectra, as represented by the \dmds distributions of Figs.~\ref{fig:ds1mass} and~\ref{fig:datafit}, the signal function consists of a relativistic Breit-Wigner line shape numerically convolved with the \pstar-dependent resolution function (Eq.~(\ref{eq:mulg})). A linear function is used to describe the background.\\
\indent The relativistic Breit-Wigner function used takes the form
\begin{equation}
\label{eq_relBWspin}
\left(\frac{p_{1,m}}{p_{1,m_{0}}}\right)^{2L+1}\left(\frac{m_{0}}{m}\right)\frac{mF_{L}(p_{1,m})^{2}}{(m_{0}^{2}-m^{2})^{2}+\Gamma_{m}^{2}m_{0}^{2}},
\end{equation}
where $m_{0}$ is an abbreviation for \dmdsz and $m$ stands for \dmds. The variable $p_{1,m}$ is the momentum of the \Dstarp in the rest frame of the \Dsop resonance candidate, which has mass $m$, and $p_{1,m_{0}}$ is the value for $m = m_{0}$. The respective Blatt-Weisskopf barrier factors $F_{L}(p_{1,m})$ for orbital angular momentum $L$ between the \Dstarp and \KS are
\begin{eqnarray}
F_{0}(p_{1,m}) & = & 1,\\
F_{1}(p_{1,m}) & = & \frac{\sqrt{1+(Rp_{1,m_{0}})^{2}}}{\sqrt{1+(Rp_{1,m})^{2}}},\\
F_{2}(p_{1,m}) & = & \frac{\sqrt{9+3(Rp_{1,m_{0}})^{2}+(Rp_{1,m_{0}})^{4}}}{\sqrt{9+3(Rp_{1,m})^{2}+(Rp_{1,m})^{4}}},
\label{eq:bwcoeff}
\end{eqnarray}
where
\begin{equation}
\label{eq_Rval}
R=1.5~(\!\gevc)^{-1}
\end{equation}
is defined as in Ref.~\cite{Hi72}. The mass-dependent width is given by
\begin{eqnarray}
\Gamma_{m} & = & \gds \Bigg( \mathcal{B}_{1}\biggl(\frac{p_{1,m}}{p_{1,m_{0}}}\biggr)^{2L+1}\biggl(\frac{m_{0}}{m}\biggr)F_{L}(p_{1,m})^{2} \nonumber \\ & + & \mathcal{B}_{2}\biggl(\frac{p_{2,m}}{p_{2,m_{0}}}\biggr)^{2L+1}\biggl(\frac{m_{0}}{m}\biggr)F_{L}(p_{2,m})^{2} \Bigg)
\label{eq:gamtot}
\end{eqnarray}
with $\gds$ the total intrinsic width of the \Dsop resonance. This relation takes into account the $\Dsop \to \Dstarp\Kz$ and the $\Dsop \to \Dstarz\Kp$ decay modes, with the corresponding branching fractions $\mathcal{B}_{1}$ and $\mathcal{B}_{2}$, respectively: 
\begin{equation}
\mathcal{B}_{i} = \frac{p_{i,m_{0}}^{2L+1}}{p_{1,m_{0}}^{2L+1} + p_{2,m_{0}}^{2L+1}}.
\end{equation}
Since the \Dsop mass lies close to threshold for both decay modes, the mass values of the decay particles make a significant difference. The momenta $p_{2,m}$ and $p_{2,m_{0}}$ correspond to $p_{1,m}$ and $p_{1,m_{0}}$, respectively, but are calculated for the $\Dstarz\Kp$ decay mode.\\
\indent It is assumed that the \Dsop has spin-parity $J^{P}=1^{+}$ and from parity conservation that the orbital angular momentum $L$ is either $0$ or $2$. The $S$-wave usually dominates in $1^{+}$ decays, so $L=0$ is chosen for the main fit and an additional $L=2$ contribution is used to estimate a systematic uncertainty. Further discussion on the $J$ and $L$ values is presented in Sec.~\ref{sec:angdist}.\\
\indent The fit to the $\dmds = m(\Dsop)-m(\Dstar)-m(\KS)$ mass difference spectrum in data~(Fig.~\ref{fig:datafit}) yields mean mass differences
\begin{eqnarray*}
\dmdsz & = & 27.231 \pm 0.020 \mevcc\quad(K4\pi),\\
\dmdsz & = & 27.205 \pm 0.018 \mevcc\quad(K6\pi),
\end{eqnarray*} 
and total width values
\begin{eqnarray*}
\gds & = & 1.000 \pm 0.049 \mev\quad(K4\pi),\\
\gds & = & 0.941 \pm 0.045 \mev\quad(K6\pi).
\end{eqnarray*}
The fitted values for the two \Dz decay modes agree within the statistical errors. The signal yield is $3704 \pm 71$ for $K4\pi$ and $4334 \pm 78$ for $K6\pi$.
 
\section{Angular Distribution}
\label{sec:angdist}

\begin{figure}
\includegraphics[width=0.3\textwidth]{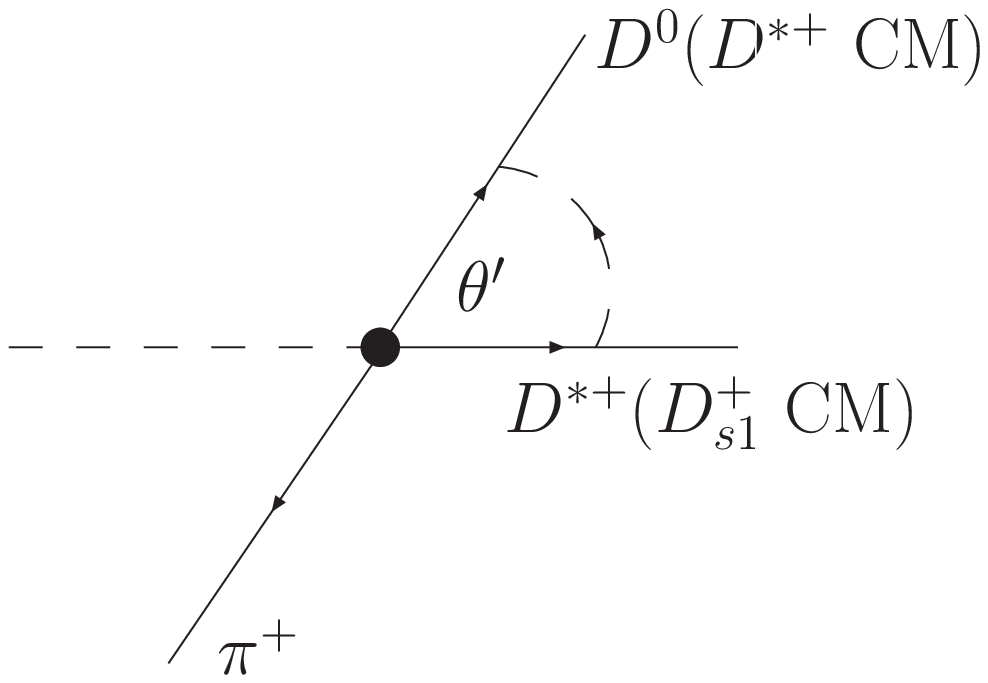}\\
\includegraphics[width=0.3\textwidth]{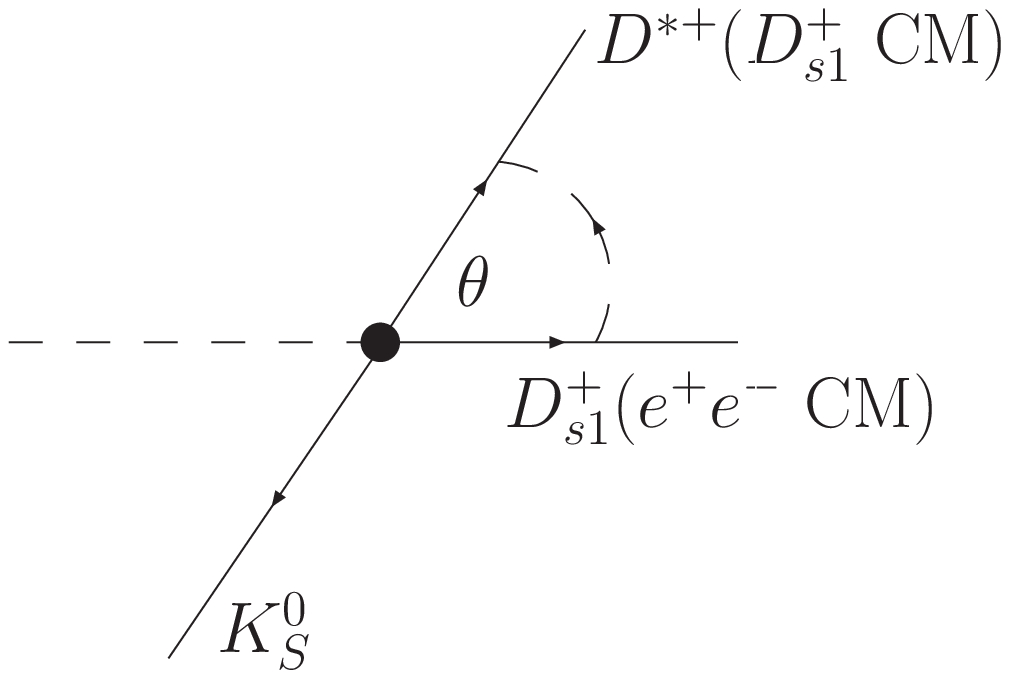}
\begin{picture}(0,0)
\put(-6.1,6.5){(a)}
\put(-6.1,3.){(b)}
\end{picture}
\caption{a) Decay angle $\theta^{\prime}$ of the \Dstarp. b) Decay angle $\theta$ of the \Dsop.}
\label{fig:ang}
\end{figure}

\begin{figure}
\includegraphics[width=0.3\textwidth]{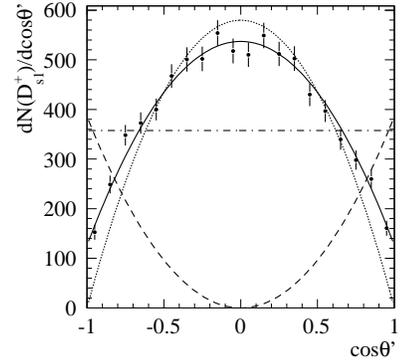}
\caption{Efficiency-corrected signal yield as function of $\cos\theta^{\prime}$ in data. The following models are fitted to the distribution: $a(\sin^{2}\theta^{\prime} + \beta\cos^{2}\theta^{\prime})$~(solid line); a constant (dash-dotted line); $a\cos^{2}\theta^{\prime}$~(dashed line); $a\sin^{2}\theta^{\prime}$~(dotted line).}
\label{fig:angdist}
\end{figure}

\begin{figure}
\includegraphics[width=0.3\textwidth]{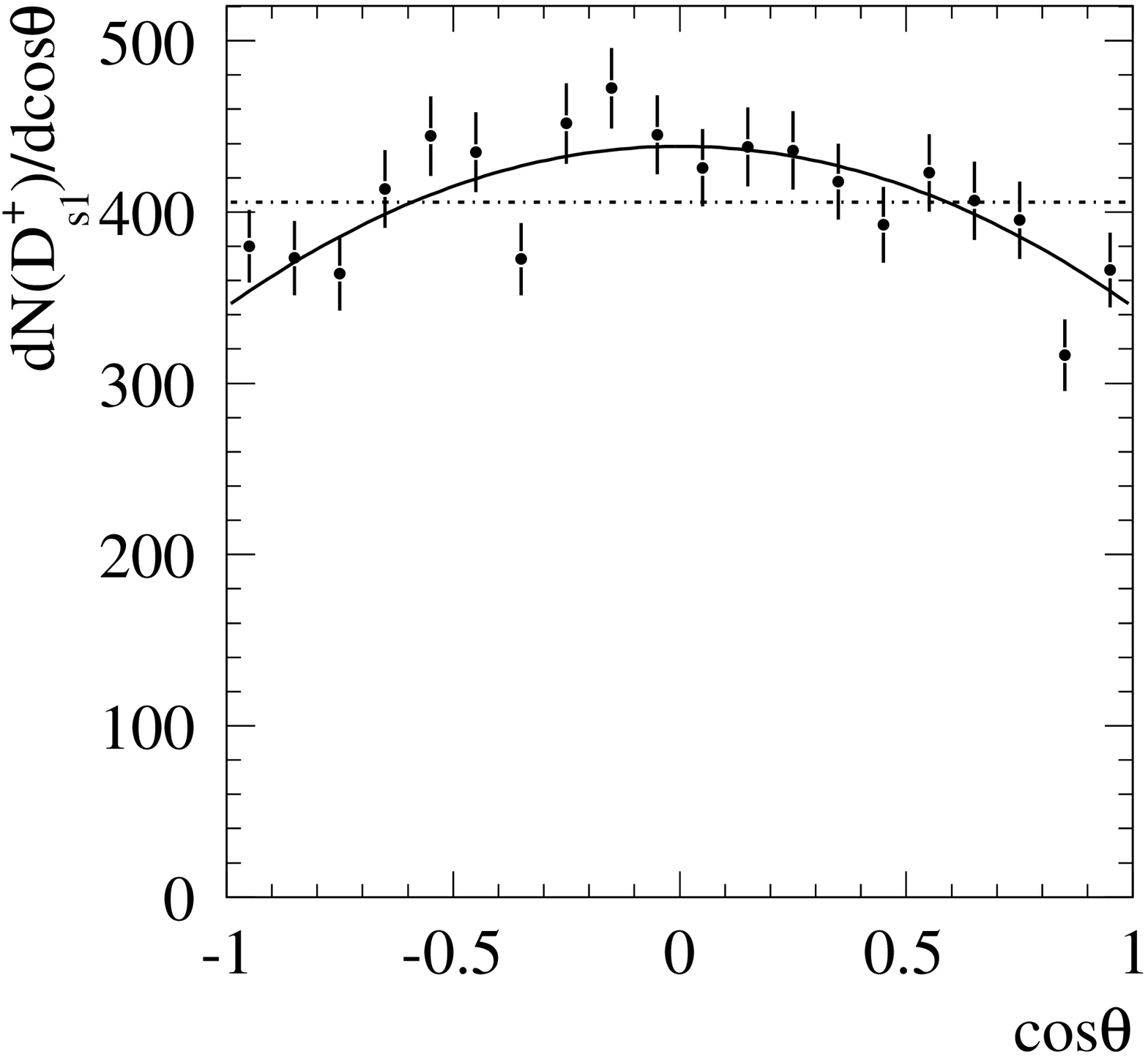}
\caption{Efficiency-corrected signal yield as function of $\cos\theta$ in data. The following models are fitted to the distribution: constant (dotted line); $a(1 + t\cos^{2}\theta)$ (solid line).}
\label{fig:angalt}
\end{figure}

The assigned spin-parity $J^{P}=1^{+}$ of the \Dsop is based on studies with small data samples (less than 200 reconstructed events)~\cite{Al93,Au08}. There, fits of an angular distribution corresponding to unnatural spin-parity ($1^{+}, 2^{-}, \ldots$) yielded the highest confidence level. In this analysis clean signals with a total number of about 8000 reconstructed \Dsop-candidates are available, making a detailed study possible.\\
\subparagraph{\boldmath \Dstarp decay angle.} Since in this analysis the origin of the \Dsop is not known, the decay angle $\theta^{\prime}$ between the \Dz momentum vector in the \Dstarp CM system and the \Dstarp momentum vector in the \Dsop CM system (Fig.~\ref{fig:ang}a) is used for the $J^{P}$ analysis. The resulting angular distribution $dN(\Dsop)/d\cos\theta^{\prime}$ is influenced by the spin of the \Dsop. The expected distributions for different \Dsop spin-parity values are calculated using the helicity formalism~\cite{Ch71, Am83, Ri84} and are listed in Table~\ref{tab:ang}.\\
\indent The data are corrected for the detection efficiency and divided into 20 bins of $\cos\theta^{\prime}$. The signal entries for the $\cos\theta^{\prime}$ bins are obtained from separate fits to the data with the mass and decay width of the \Dsop fixed to the values reported in Sec.~\ref{sec:datafit}. The $dN(D_{s1}^{+})/d\cos\theta^{\prime}$ distribution shown in Fig.~\ref{fig:angdist} is the combined result from the $K4\pi$ and $K6\pi$ samples.\\ 
\indent Comparison with the theoretical distributions shows a clear preference for the unnatural spin-parity values $J^{P} = 1^{+},2^{-},3^{+}\ldots$, confirming the earlier results~\cite{Au08,Al93}. The signal function for these $J^{P}$ values is
\begin{equation}
I(\theta^{\prime}) = a(\sin^{2}\theta^{\prime} + \beta\cos^{2}\theta^{\prime}),
\label{eq:ithetap}
\end{equation} 
where $\beta = |A_{00}|^{2}/|A_{10}|^{2}$ and $a$ is a constant. The helicity amplitudes $|A_{00}|$ and $|A_{10}|$ correspond to the \Dstarp helicities $0$ and $\pm1$, respectively. \\
\indent The lowest value $J^{P}=1^{+}$ is the most probable one: assuming $1^{+}$ implies $l = 1$ (orbital momentum between the light and heavy quark), while the higher $J$ values demand $l\geq2$; such mesons are expected to be highly suppressed in $\epem$ production~\cite{Fi83}. The coefficient $\beta$ is 1 in the case of a pure $S$-wave decay of \Dsop into $\Dstarp\KS$, thus yielding a flat distribution in disagreement with data. The results reported in Table~\ref{tab:ang} for $\beta$ clearly indicate a $D$-wave contribution. Based on the results for $\beta$, the ratio of the helicity amplitudes is determined to be $|A_{10}|/|A_{00}|=2.09\pm0.09$ for the combined $K4\pi$ and $K6\pi$ samples, and $2.09 \pm 0.14$ and $2.04\pm0.13$ for the individual samples, respectively. The squared ratio of the amplitudes is $|A_{10}|^2/|A_{00}|^2 = 4.35 \pm 0.38$ (combined data), consistent with the Belle result $|A_{10}|^2/|A_{00}|^2={3.6\pm0.3\pm0.1}$~\cite{Ba08}.\\
\subparagraph{\boldmath \Dsop decay angle.} The $dN(D_{s1}^{+})/d\cos\theta$ distribution is also studied, where $\theta$ is the decay angle between the \Dstarp momentum vector in the \Dsop CM system and the \Dsop momentum vector in the \epem CM system (Fig.~\ref{fig:ang}b). The combined efficiency-corrected $\cos\theta$ spectrum is shown in Fig.~\ref{fig:angalt}. The results in this figure indicate that the \Dsop decay to $\Dstarp\KS$ is not purely $S$-wave. Were this decay purely $S$-wave, the distribution would be flat. The $\cos\theta$ distribution, assuming $J^{P}=1^{+}$, is
\begin{eqnarray}
I(\theta) & = & a((1 + \rho_{00})|A_{10}|^{2}+(1-\rho_{00})|A_{00}|^{2} \nonumber\\
& + & (1-3\rho_{00})(|A_{10}|^{2}-|A_{00}|^{2})\cos^{2}\theta)
\label{eq:itheta}
\end{eqnarray}
where $\rho_{00}$ gives the probability that the \Dsop helicity is zero. \\
\indent Results from a fit of both a constant and a distribution proportional to $1 + t\cos^{2}\theta$ (based on Eq.~(\ref{eq:itheta})) are given in Table~\ref{tab:angalt}. Using the value of $t$ from the $\cos\theta$ fit, the result for $|A_{10}|^{2}/|A_{00}|^{2}$ from the $\cos\theta^{\prime}$ fit, and the coefficients from Eq.~(\ref{eq:itheta}), we determine $\rho_{00}=0.48\pm0.03$ for the combined $K4\pi$ and $K6\pi$ samples, and $0.44 \pm 0.04$ and $0.53 \pm 0.04$ for the individual samples, consistent with the Belle result $\rho_{00}=0.490\pm 0.012$~\cite{Ba08}.\\
\indent Several effects that might affect the results of the angular analysis have been studied. 
\subparagraph{Test for non-flat efficiency.} The formalism used for the calculation of $I(\theta^{\prime})$ assumes a flat acceptance in $\cos\theta$. In this study, the efficiency decreases a few percent for values of $\cos\theta > 0$. In order to assess the impact of this effect, all \Dsop candidates with $\cos\theta>0$ are removed from the data sample. The results for $\beta$ from fits to the reduced $\cos\theta^{\prime}$ spectra are consistent with the nominal results, ruling out an observable effect due to non-flat efficiency. 
\subparagraph{Test for possible interference.} Possible interference with unreconstructed recoil particle(s) $X$ in the decay chain $\epem \to \Dsop X$ is considered. The effect of interference is expected to depend on the flight direction of the \Dsop. Therefore the data are divided into four sub-samples based on their $\cos\theta_d$ value, where $\theta_d$ is the flight angle of the \Dsop relative to the beam axis (calculated in the \epem CM system). For each of these reduced data samples, the fit to the $\cos\theta^{\prime}$ distribution is repeated. The values obtained for the parameter $\beta$ are fully consistent within errors with each other and with the nominal value (full data sample), ruling out a significant interference effect. The same consistency between results is found in fits to $\cos\theta$.

\begin{table*}
\caption{List of spin-parity values $J^{P}$ for the \Dsop and the corresponding decay angle distributions for the \Dstarp. Under the assumption of a strong decay, $0^{+}$ is forbidden. The last three columns show the $\chi^{2}/N\!D\!F$ of the fits to the $\cos\theta^{\prime}$-distribution for efficiency-corrected data, with $N\!D\!F$ being the number of degrees of freedom.}
\begin{ruledtabular}
  \begin{tabular}{ccccc}
    $J^{P}$ &  $dN(D_{s1}^{+})/d\cos\theta^{\prime}$ & $\chi^{2}/N\!D\!F (K4\pi)$ & $\chi^{2}/N\!D\!F (K6\pi)$ & $\chi^{2}/N\!D\!F$ (combined data) \\ \hline 
   $0^{+}$ & forbidden & $-$ & $-$ & $-$\\ 
   $0^{-}$ & $a\cos^{2}\theta^{\prime}$ & $2142.7/19$ & $2440.8/19$ & $4578.0/19$ \\ 
   $1^{-}, 2^{+}, 3^{-}, \ldots$ & $a\sin^{2}\theta^{\prime}$ & $103.2/19$ & $108.8/19$ & $190.9/19$ \\ 
    $1^{+}, 2^{-}, 3^{+}, \ldots$ ($S$-wave only)& const & $392.1/19$ & $425.1/19$ & $802.5/19$\\
   $1^{+}, 2^{-}, 3^{+}, \ldots$ ($S$-, $D$-wave) & $a(\sin^{2}\theta^{\prime} + \beta\cos^{2}\theta^{\prime})$ & $24.9/18$ & $9.5/18$ & $14.7/18$ \\ 
 & & $(\beta=0.23\pm0.03)$ & $(\beta=0.24\pm0.03)$ & $(\beta=0.23\pm0.02)$ \\
  \end{tabular}
\label{tab:ang}
\end{ruledtabular}
\end{table*}

\begin{table*}
\caption{$\chi^{2}/N\!D\!F$ values of the fits to the $\cos\theta$-distribution for efficiency-corrected data, with $N\!D\!F$ being the number of degrees of freedom.}
\begin{ruledtabular}
  \begin{tabular}{ccccc}
     &  $dN(D_{s1}^{+})/d\cos\theta$ & $\chi^{2}/N\!D\!F (K4\pi)$ & $\chi^{2}/N\!D\!F (K6\pi)$ & $\chi^{2}/N\!D\!F$ (combined data) \\ \hline 
    pure $S$-wave & constant & $19.0/19$ & $55.5/19$ & $57.0/19$ \\ 
    $S$- and $D$-wave & $a(1 + t\cos^{2}\theta)$ & $12.0/18$ & $27.3/18$ & $25.2/18$ \\ 
   & & $(t=-0.15\pm0.05)$ & $(t=-0.27\pm0.05)$ & $(t=-0.21\pm0.04)$ \\
  \end{tabular}
\label{tab:angalt}
\end{ruledtabular}
\end{table*}

\section{Systematic studies}
\label{sec:syst}
The investigated sources of systematic uncertainty can be separated into three main categories: uncertainties arising from the resolution model, fit procedure, and reconstruction. 
The uncertainties are defined by taking the differences \delm and \delg between the standard result for the mass difference \dmdsz and width \gds given in Sec.~\ref{sec:datafit} and the result obtained with the correspondent modification. A summary of the results is listed in Table~\ref{tab:systcomb}. If not otherwise stated, the momentum-dependent resolution model and the relativistic Breit-Wigner signal function combined with a first order polynomial for background parametrization from the standard fit are used. Deviations smaller than $0.5 \kevcc$ for \delm and smaller than $0.5 \kev$ for \delg are considered as negligible.  

\subsection{Resolution model uncertainties}
\label{sec:systa}
\subparagraph{General comparison between MC and Data.}
The \Dz and \KS test samples (see Sec.~\ref{sec:mc}) demonstrate that the mass resolution is underestimated by $10 \%$ in MC (Fig.~\ref{fig:resrat}), yielding an overestimated decay width from the fits to data. The effect of this is quantitatively studied by increasing the width of the resolution function by $10 \%$. The repeated fits yield no significant deviations for the mass difference, but a $51~(45)\kev$ smaller decay width. The nominal decay width values obtained from the fits in Sec.~\ref{sec:datafit} are thus corrected by these values, yielding values of $\gds= 0.949~(0.896)~\mev$ for the $K4\pi$ ($K6\pi$) mode.\\
\indent To estimate the corresponding systematic uncertainty, the resolution function modification is varied within $(10 \pm 4)\%$ to take a possible $p^{\ast}$ dependence into account (this value is derived from Fig.~\ref{fig:resrat}(a), which shows the largest variation in $p^{\ast}$). There are no effects on $\dmdsz$, and a deviation of $^{-30}_{+8} \kev$ for \gds is observed in both decay modes, compared to the corrected results from above. As a conservative estimate, the larger deviation is used as a two-sided uncertainty, providing the largest systematic error for the decay width. 

\subparagraph{Further validation of the resolution model.}
To further validate the procedure used to obtain the resolution model, the results from fits to the $\Gamma_{1}$ and $\Gamma_2$ MC samples are compared. The derived resolution function parameters are in good agreement between the two samples. The widths of the reconstructed \dmds distributions determined from fitting the $\Gamma_{2}$ samples, $\gds = 2.004 \pm 0.016 \mev$ for the $K4\pi$ mode and $2.018\pm0.022 \mev$ for the $K6\pi$ mode, are in good agreement with the generated values. Similarly, when the resolution function from the $\Gamma_{2}$ sample is used to determine the width for the $\Gamma_{1}$ sample, values of $\gds = 1.003 \pm 0.005\mev$ and $0.999\pm0.006 \mev$, respectively, are obtained, in agreement with the generated values.\\
\indent As a conservative estimate, the small deviations found during the validation procedure for the resolution model using the $\Gamma_{1}$ sample in Sec.~\ref{sec:res} are used as systematic uncertainties: $\delm = -7~(-10) \kevcc$; $\delg = +1~(-9) \kev$ for $K4\pi~(K6\pi)$.  

\subparagraph{Alternative resolution models.}
Using the resolution model obtained from the $\Gamma_{2}$ MC sample, a fit to data yields uncertainties $\delm < 0.5~(<0.5) \kevcc$ and $\delg = +1~(+12) \kev$ for $K4\pi$~($K6\pi$).\\
\indent Instead of the momentum-dependent resolution model of the standard analysis, an alternative model has been tested, based on the comparison of MC and data distributions that show disagreement, such as the $\pstar$ dependence of the \Dsop yield. By dividing the MC and data spectra from Fig.~\ref{fig:pstardiff}, a correction function is derived. MC data are modified with this function such that the two distributions in Fig.~\ref{fig:pstardiff} coincide. From these corrected MC, a new resolution model is derived. The results for $\dmdsz$ and $\gds$ in data agree within the error with the momentum-dependent treatment (systematic uncertainties $\delm < 0.5~(<0.5) \kevcc, \delg = -2~(+1) \kev$ for $K4\pi$~($K6\pi$)).\\
\indent The larger deviations listed above are reported as the systematic uncertainties associated with the use of alternative resolution models.  

\subparagraph{Parameters of the {\boldmath $\pstar$}-dependent resolution model.}
The parameter $r$ of the $\pstar$-dependent resolution model is modified within its error leading to negligible deviations in $\dmdsz$ and $\pm 6~(\pm 7)\kev$ in $\gds$ for $K4\pi$~($K6\pi$). A different parametrization of the $\sigma_{0}(\pstar)$-dependence (second order polynomial) results in a negligible deviation for $\dmdsz$ and $-3~(-2)$ \kev for \gds. 

\subsection{Fit Procedure Uncertainties}

\subparagraph{Breit-Wigner line shape.}
In the standard fit, a pure $S$-wave decay of the \Dsop to $\Dstarp\KS$ is assumed, using a Breit-Wigner line shape corresponding to $L=0$. To estimate a systematic uncertainty, a combination of an $S$-wave and a $D$-wave Breit-Wigner is used instead. Relative contributions of $72\%$ and $28\%$ are used, based on a decay angle analysis of the \Dsop by the Belle Collaboration~\cite{Ba08}. Using the modified signal lineshape, uncertainties of $-9~(-8)\kevcc$ in $\dmdsz$ and $-2~(-3)\kev$ in $\gds$ are derived, compared with the standard results.\\
\indent As an additional check, the value of $R$ (Eq.~(\ref{eq_Rval})) is set to $2.0~(\!\gevc)^{-1}$. No effect on $\dmdsz$ and \gds is observed.\\
\indent The effect of neglecting $\Dstarz\Kp$ decays (Sec.~\ref{sec:datafit}) is studied by setting $\mathcal{B}_{1}=1$ and $\mathcal{B}_{2}=0$. The resulting uncertainties are negligible for both $\dmdsz$ and $\gds$. 
 
\subparagraph{Numerical precision of convolution.}
The integration range and number of steps in the numerical convolution of the signal line shape and resolution function (Sec.~\ref{sec:datafit}) are varied, resulting in a negligible deviation both for the mass and the width.

\subparagraph{Mass window.} 
The mass window for \dmds is enlarged, resulting in no significant change for $\dmdsz$ and a difference for the width of $\delg = +9~(+3)\kev$. 

\subparagraph{Background parametrization.} 
For background parametrization, a power law function proportional to $m^{\alpha}$ is used instead of a linear function, leaving $\dmdsz$ unaffected but yielding $\delg = -5~(-7)\kev$ for $K4\pi~(K6\pi)$. 

\subsection{Reconstruction Uncertainties}

\subparagraph{Tracking region material.}
Uncertainties in the \Dsop mass may arise from uncertainties in the energy-loss correction in charged particle tracking. Studies of $\Lambda$ and \KS decays suggest that the correction might be underestimated~\cite{Au05}. Two possibilities are considered, one with the SVT material density increased by $20\%$ and the other with the tracking region material density (SVT, DCH) increased by $10\%$, as was investigated in detail in Refs.~\cite{Au05,Au06}. The deviations indicate that the fit results for the mass might be underestimated. The larger values from the two studies ($\delm = +21~(+13)\kevcc$ and $\delg = +14~(-15)\kev$  for $K4\pi~(K6\pi)$) are chosen as a two-sided systematic uncertainty.  

\subparagraph{SVT alignment.}
Slight possible deviations in the alignment of SVT components may affect the measurement of angles between tracks and thus the mass measurement. This is studied by applying small distortions to the SVT alignment in simulated data, comprising general changes between different run periods and radial shifts. Results are $\delm = \pm 6~(\pm7)\kevcc$ and $\delg = \pm 2~(\pm14)\kev$  for $K4\pi~(K6\pi)$. 

\subparagraph{Magnetic field.}
The magnetic field inside the tracking volume has several components: the main solenoidal field, fields from permanent magnets and an additional magnetization of the latter due to the main field. To understand the effect of the field on the track reconstruction, the solenoid field strength is varied by $\pm 0.02 \%$ and the magnetization of the permanent magnets by $\pm 20 \%$~\cite{Au05,Au06}. For the mass difference, the larger deviations arising from the change in rescaled solenoid field and magnetization are added in quadrature and the sum is assigned as a systematic uncertainty associated with the magnetic field; the same is done for the decay width. The results are $\delm = \pm 13~(\pm19)\kevcc$ and $\delg = \pm 19~(\pm11)\kev$ for $K4\pi~(K6\pi)$. 

\subparagraph{Distance scale.}
A further source of uncertainty for the momentum determination comes from the distance scale. The positions of the signal wires in the drift chamber are known with a precision of $40\mum$, corresponding to a relative precision of $0.01\%$. As an estimate of the uncertainty of the momentum due to the distance scale, a systematic error half the size of the uncertainty obtained from the $0.02\%$ variation of the solenoid field is assigned. For the mass difference this yields a shift of $\pm4~(\pm6) \kevcc$ for $K4\pi~(K6\pi)$; the width is shifted by $\pm8~(\pm4) \kev$ for $K4\pi~(K6\pi)$.

\subparagraph{Drift Chamber hits.}
In the standard \Dsop selection no lower limit is set for the number of drift chamber hits. Requiring at least 20 hits per track, thereby excluding the low momentum pions from \Dstarp decays, modifies $\dmdsz$ by $-11~(-15) \kevcc$ and \gds by $-7~(-7) \kev$ for $K4\pi~(K6\pi)$.
 
\subparagraph{Angular dependence.}
For the reconstructed \KS and \Dz masses from the test data samples (see Sec.~\ref{sec:mc}), a sine-like dependence on the azimuthal angle $\phi$ is observed. This effect was also observed in a previous \babar~analysis and might be related to the internal alignment of the DCH~\cite{Au05}. For a detailed study, the same $\phi$-dependence is introduced into the signal MC samples by modifying the kaon and pion track momenta accordingly. Due to symmetry, the effect disappears when all $\phi$ angles are taken into account, but as a conservative estimate the amplitude of the sine-like shift on the reconstructed \Dsop mass in MC ($13~(14)\kevcc$ for $K4\pi~(K6\pi)$) is taken as a systematic error for $\dmdsz$. 

\section{Results}
\label{sec:results}
For the combination of the measurements, a Best Linear Unbiased Estimate (BLUE,~\cite{Ly88}) technique is used, where correlations between the systematic uncertainties are taken into account. Adding the nominal \Dstarp and \KS masses, $2010.25\mevcc$ and $497.614\mevcc$ (with their respective errors of $0.140\mevcc$ and $0.024\mevcc$~\cite{Pd10}), the final value for the \Dsop mass is
\begin{equation*}
m(\Dsop) = 2535.08 \pm 0.15 \mevcc.
\end{equation*}
Using a slightly different method for the combination of the individual results~\cite{Au06}, a value of
\begin{equation*}
m(\Dsop) = 2535.08 \pm 0.01 \pm 0.15 \mevcc
\end{equation*}
is obtained, where the first error denotes the statistical and the second the systematic uncertainty. The latter is dominated by the uncertainty of the \Dstarp mass. The mass difference between the \Dsop and the \Dstarp is
\begin{equation*}
m(\Dsop) - m(\Dstarp) = 524.83 \pm 0.04 \mevcc,
\end{equation*}
using the BLUE technique, and for the alternative combination method
\begin{equation*}
m(\Dsop) - m(\Dstarp) = 524.83 \pm 0.01 \pm 0.04 \mevcc,
\end{equation*}
which has a significantly smaller systematic uncertainty than the $m(\Dsop)$ result.\\
\indent For the total decay width of the \Dsop, combining the results from the two measurements in the same way as for the mass yields
\begin{equation*}
\gds = 0.92 \pm 0.05 \mev,
\end{equation*}
using the BLUE technique, and for the alternative combination method
\begin{equation*}
\gds = 0.92 \pm 0.03 \pm 0.04 \mev.
\end{equation*}
The corrections of $-51~(-45)\kev$ for the $K4\pi$ ($K6\pi$) decay mode, based on the systematic resolution studies (Sec.~\ref{sec:systa}), are applied prior to the combination process.

\begin{table*}
\caption{Summary of the systematic uncertainties for the mass difference (\delm) and for the decay width (\delg).}
\begin{ruledtabular}
\begin{tabular}{lcccc} 
& \multicolumn{2}{c}{\delm / \kevcc} & \multicolumn{2}{c}{\delg / \kev} \\ 
Systematic uncertainty & $K4\pi$ & $K6\pi$ & $K4\pi$ & $K6\pi$ \\ \hline
Resolution $+10~\%$ & $<0.5$ & $<0.5$ & $\pm30$ & $\pm30$  \\  
MC validation & $\pm 7$ & $\pm 10$ & $\pm 1$ & $\pm 9$  \\ 
Alternative resolution models & $<0.5$ & $<0.5$ & $\pm 2$ & $\pm 12$  \\ 
Multi-Gaussian resolution: $r \pm \delta r$ & $<0.5$ & $<0.5$ & $\pm 6$ & $\pm 7$  \\ 
Multi-Gaussian resolution: Parametrization of $\sigma_{0}$ & $<0.5$ & $<0.5$ & $\pm 3$ & $\pm2$  \\ 
Breit-Wigner signal line shape: Value of $L$ & $\pm 9$ & $\pm 8$ & $\pm 2$ & $\pm 3$ \\ 
Mass window for \dmds & $<0.5$ & $<0.5$ & $\pm 9$ & $\pm 3$  \\ 
Background parametrization & $<0.5$ & $<0.5$ & $\pm 5$ & $\pm 7$ \\ 
Tracking region material density & $\pm 21$ & $\pm 13$ & $\pm 14$ & $\pm 15$ \\ 
SVT Alignment & $\pm6$ & $\pm7$ & $\pm2$ & $\pm14$ \\ 
Magnetic field strength & $\pm 13$ & $\pm 19$ & $\pm 19$ & $\pm 11$  \\
Length scale & $\pm4$ & $\pm6$ & $\pm8$ & $\pm4$  \\ 
Drift chamber hits & $\pm 11$ & $\pm 15$ & $\pm 7$ & $\pm 7$  \\ 
$\phi$-dependency & $\pm13$ & $\pm14$ & $\cdots$ & $\cdots$  \\ 
\end{tabular}
\label{tab:systcomb}
\end{ruledtabular}
\end{table*}

\section{Summary}
\label{sec:summ}
In this paper, precision measurements of the mass and decay width of the charmed-strange meson \Dsopnum via the decay $\Dsop \rightarrow \Dstarp \KS$ are presented. Two decay modes are analyzed, with the \Dz that originates from the \Dstarp decaying either through $\Km\pip$ or $\Km\pip\pip\pim$.\\
\indent The results include the first significant measurement of the total decay width of the \Dsop. This width is determined to be
\begin{equation*}
\gds = 0.92 \pm 0.03 \pm 0.04 \mev,
\end{equation*}
compared to the $90\%$ confidence level upper limit of $2.3 \mev$ given in Ref.~\cite{Pd10}.
The mass of the \Dsopnum is measured to be
\begin{equation*}
m(\Dsop) = 2535.08\pm 0.01 \pm 0.15 \mevcc,
\end{equation*} 
and the $\Dsop-\Dstarp$ mass difference to be
\begin{equation*}
m(\Dsop)- m(\Dstarp) = 524.83 \pm 0.01 \pm 0.04 \mevcc.
\end{equation*}
The result for the $\Dsop-\Dstarp$ mass difference represents a significant improvement compared to the current world average of $525.04 \pm 0.22 \mevcc$~\cite{Pd10}. \\
\indent Based on a decay angle analysis, the $J^{P} = 1^{+}$ assignment for the \Dsop meson is confirmed. 

\begin{acknowledgments}
We are grateful for the
extraordinary contributions of our \pep2\ colleagues in
achieving the excellent luminosity and machine conditions
that have made this work possible.
The success of this project also relies critically on the
expertise and dedication of the computing organizations that
support \babar.
The collaborating institutions wish to thank
SLAC for its support and the kind hospitality extended to them.
This work is supported by the
US Department of Energy
and National Science Foundation, the
Natural Sciences and Engineering Research Council (Canada),
the Commissariat \`a l'Energie Atomique and
Institut National de Physique Nucl\'eaire et de Physique des Particules
(France), the
Bundesministerium f\"ur Bildung und Forschung and
Deutsche Forschungsgemeinschaft
(Germany), the
Istituto Nazionale di Fisica Nucleare (Italy),
the Foundation for Fundamental Research on Matter (The Netherlands),
the Research Council of Norway, the
Ministry of Education and Science of the Russian Federation,
Ministerio de Educaci\'on y Ciencia (Spain), and the
Science and Technology Facilities Council (United Kingdom).
Individuals have received support from
the Marie-Curie IEF program (European Union) and
the A. P. Sloan Foundation.
\end{acknowledgments}


\begin{thebibliography}{99}
\bibitem{Au03} B.~Aubert {\it et al.} (\babar~Collaboration), \jprl{90}, 242001 (2003).
\bibitem{Au04} B.~Aubert {\it et al.} (\babar~Collaboration), \jprl{93}, 181801 (2004).
\bibitem{Mi04} V.~Mikani {\it et al.} (Belle Collaboration), \jprl{92}, 012002 (2004).
\bibitem{Au06} B.~Aubert {\it et al.} (\babar~Collaboration), \jprd{74}, 032007 (2006).
\bibitem{Be03} D.~Benson {\it et al.} (CLEO Collaboration), \jprd{68}, 032002 (2003).
\bibitem{Kr03} P.~Krokovny {\it et al.} (Belle Collaboration), \jprl{91}, 262002 (2003).
\bibitem{Ca03} R.N.~Cahn and J.D.~Jackson, \jprd{68}, 037502 (2003).
\bibitem{Ba03} T.~Barnes, F.E.~Close, and H.J.~Lipkin, \jprd{68}, 054006 (2003).
\bibitem{Bd03} W.A.~Bardeen, E.J.~Eichten, and C.T.~Hill, \jprd{68}, 054024 (2003).
\bibitem{No03} M.A.~Nowak, M.~Rho, and I.~Zahed, \app{B35}, 2377 (2004).
\bibitem{Bv03} E.~van~Beveren and G.~Rupp, \jprl{91}, 012003 (2003).
\bibitem{Ko04} E.E.~Kolomeitsev and M.F.M.~Lutz, \plb{582}, 39 (2004).
\bibitem{Ma05} L.~Maiani, F.~Piccinini, A.D.~Polosa, and V.~Riquer, \jprd{71}, 014028 (2005).
\bibitem{Te03} K.~Terasaki, \jprd{68}, 011501 (2003).
\bibitem{Do03} A.~Dougall, R.D.~Kenway, C.M.~Maynard and C.~McNeile (UKQCD Collaboration), \plb{569}, 41 (2003).
\bibitem{Bl03} G.S.~Bali, \jprd{68}, 071501 (2003).
\bibitem{Co04} P.~Colangelo, F.~de~Fazio, and R.~Ferrandes, \mpl{A19}, 2083 (2004).
\bibitem{Sw06} E.S.~Swanson, \prep{429}, 243 (2006).
\bibitem{CC} The use of charge conjugated reactions is implied throughout the text.
\bibitem{Al89} H.~Albrecht {\it et al.} (ARGUS Collaboration), \plb{230}, 162  (1989).
\bibitem{Pd10} K.~Nakamura {\it et al.} (Particle Data Group), \jpg{37}, 075021 (2010).
\bibitem{Al93} J.~Alexander {\it et al.} (CLEO Collaboration), \plb{303}, 377 (1993).
\bibitem{Au08} B.~Aubert {\it et al.} (\babar~Collaboration), \jprd{77}, 011102(R) (2008).
\bibitem{Ba08} V.~Balagura {\it et al.} (Belle Collaboration), \jprd{77}, 032001 (2008).
\bibitem{Au02} B.~Aubert {\it et al.} (\babar~Collaboration), \nima{479}, 1 (2002).
\bibitem{La01} D.~Lange, \nima{462}, 152 (2001).
\bibitem{Ag03} S.~Agostinelli {\it et al.} (GEANT4 Collaboration), \nima{506}, 250 (2003).
\bibitem{Hi72} F.~von~Hippel and C.~Quigg, \jprd{5}, 624 (1972).
\bibitem{Ch71} S.~Chung, CERN Yellow Report 71-8 (1971).
\bibitem{Am83} C.~Amsler and C.~Bizot, \cpc{30}, 21 (1983).
\bibitem{Ri84} J.~Richman, CALT-68-1148 (1984).
\bibitem{Fi83} R.D.~Field and S.~Wolfram, \npb{213}, 65 (1983).
\bibitem{Au05} B.~Aubert {\it et al.} (\babar~Collaboration), \jprd{72}, 052006 (2005).
\bibitem{Ly88} L.~Lyons, D.~Gibaut, and P.~Clifford, \nima{270}, 110 (1988).
\end{thebibliography}
\end{document}